\documentclass[12pt,a4paper]{article}

\usepackage{epsfig}
\usepackage{amsmath,amsfonts,amssymb}
\usepackage{t1enc}
\usepackage{cite}
\usepackage{colortbl}
\usepackage{pifont}

\providecommand{\openone}{\leavevmode\hbox{\small1\kern-3.8pt\normalsize1}}

\newcommand{\bmu}{\mathcal{B}_\mu}

\newcommand{\gmu}{\mathcal{G}_\mu}

\newcommand{\omf}{\omega^4}
\newcommand{\OMf}{\Omega^4}

\newcommand{\afb}{A_\text{FB}}
\newcommand{\ac}{A_\text{C}}

\newcommand{\oh}{\textstyle \frac{1}{2}}

\newcommand{\gM}{\gamma^\mu}

\newcommand{\la}{\lambda^a}

\begin{document}

\begin{center}
\begin{Large}
{\bf Simple models for the top asymmetry: \\[2mm]
constraints and predictions }
\end{Large}

\vspace{0.5cm}
J. A. Aguilar--Saavedra, M. P\'erez-Victoria \\[0.2cm] 
{\it Departamento de F\'{\i}sica Te\'orica y del Cosmos and CAFPE, \\
Universidad de Granada, E-18071 Granada, Spain}
\end{center}

\begin{abstract}
We perform a comprehensive study of the allowed range for the Tevatron $t \bar t$ forward-backward asymmetries in six representative new physics models: a flavour-changing $Z'$ boson, a  scalar isodoublet, a $W'$ boson, a heavy axigluon, a colour-triplet and a colour-sextet scalar. We devote special attention to the constraints from the $t \bar t$ tail at LHC on the parameter space, which will be dramatic if the measurements agree with the Standard Model prediction, specially for $Z'$ and $W'$ bosons.  We also study the predictions for the charge asymmetries at LHC and compare several proposed definitions.
\end{abstract}

\section{Introduction}

After the discovery of the top quark~\cite{Abe:1995hr,Abachi:1995iq}, a large number of events with $t \bar t$ pairs has been accumulated at the Fermilab Tevatron. A continuing discrepancy between the experimental measurements and the standard model (SM) prediction for the forward-backward (FB) asymmetry is observed in the data, both from the D0~\cite{Abazov:2007qb} and CDF~\cite{Aaltonen:2008hc,Aaltonen:2011kc} collaborations. Moreover, CDF has reported larger deviations for high $t \bar t$ invariant masses and for high rapidity differences. This situation has motivated a number of possible explanations that invoke physics beyond the SM. Since $t \bar t$ production at Tevatron energies is dominated by $q \bar q$ annihilation, $q=u,d$, it is likely that the new effects, if present, take place in $q \bar q \to t \bar t$ processes. Furthermore, because the discrepancies are large, it is plausible that they are due to tree-level exchange of new particles.

All the possible extra vector bosons and scalars that contribute to $q \bar q \to t \bar t$ have been classified in Ref.~\cite{AguilarSaavedra:2011vw} into irreducible representations of the full $SU(3)_C\times SU(2)_L \times U(1)_Y$ gauge group\footnote{This classification does not include fields that could only contribute to this process via mixing with the SM fields, after electroweak symmetry breaking. Such possibility is highly constrained by many other observables, which leaves little room for the required big effect.}. These representations constitute a `basis' in which arbitrary new physics contributions at the tree level, in the form of  heavy or light particles, can be expanded---much in the same way as a basis of gauge-invariant effective operators can be used to parameterise new physics corrections arising from a high scale. In practice, when performing explicit analyses, one needs to choose particular `directions' in the general multi-dimensional space of new physics models. Hence, although the formalism is completely general, most of the proposed models in the literature
actually correspond to one extra particle belonging to an irreducible representation. In this sense, these models can be considered minimal or, as we will refer to in the following, `simple'. The simple models which have received most attention are: (i) a new colour-octet produced in the $s$ channel~\cite{Ferrario:2008wm,Frampton:2009rk,Djouadi:2009nb,Delaunay:2010dw, Burdman:2010gr,Alvarez:2010js,Bai:2011ed,Barcelo:2011fw,Haisch:2011up}; (ii) a flavour-violating $Z'$~\cite{Jung:2009jz,Cao:2011ew,Bhattacherjee:2011nr,Berger:2011ua} or $W'$ boson~\cite{Cheung:2009ch,Cao:2010zb,Shelton:2011hq} exchanged in the $t$ channel; (iii) a charge $4/3$ scalar exchanged in the $u$ channel~\cite{Shu:2009xf,Arhrib:2009hu,Dorsner:2009mq,Patel:2011eh,Ligeti:2011vt,Grinstein:2011yv}. Some comparative studies between them have been performed, too~\cite{Cao:2009uz,Choudhury:2010cd,Gresham:2011pa}. In addition, there exist several analyses in terms of effective operators, which are valid only for heavy new physics~\cite{Jung:2009pi,AguilarSaavedra:2010zi,Zhang:2010dr,Degrande:2010kt,Blum:2011up,AguilarSaavedra:2011vw}. The effective Lagrangian can, under the assumption of decoupling new physics, be derived from the direct description in terms of extra fields, as done explicitly in Ref.~\cite{AguilarSaavedra:2011vw}. 

Thanks to the fresh data quickly flowing from the CERN Large Hadron Collider (LHC), these proposals will soon receive independent, often decisive tests. For example, the CMS Collaboration has recently presented the first measurement of a charge asymmetry in $t \bar t$ production at LHC~\cite{afbCMS}, 
\begin{equation}
\ac = 0.060 \pm 0.134\;\text{(stat)} \pm 0.026\;\text{(syst)}
\label{ec:acX}
\end{equation}
The present number has a large statistical uncertainty, but a far better precision will soon be reached, while systematic uncertainties are also expected to improve with a better knowledge of the detector. On the other hand, the $t \bar t$ invariant mass spectrum is being measured with increasing precision at the LHC energies~\cite{CMStail}. For example, with the luminosity of 1 fb$^{-1}$ already collected, around 180 semileptonic $t \bar t$ events (up to detection efficiencies) are expected at the tail with $m_{t \bar t} > 1$ TeV, so the cross section measurement will be dominated by systematics soon. The tail of the invariant mass distribution is expected to be especially sensitive to new physics in $q\bar{q}\to t\bar{t}$, since the $q$ and $\bar{q}$ parton distribution functions become more important than the gluonic one for $x\geq 0.15$. 
There are various predictions of new signals at the LHC in other processes, expected in the different models that explain the excess in the FB asymmetry~\cite{Gresham:2011dg,Hewett:2011wz}. However, in most cases the nonobservation of those signals would not rule out a given scenario, as a null result can usually be accomodated with suitable adjustments. For instance, $t$-channel exchange of a real $Z'$ boson leads to like-sign top pair production~\cite{Jung:2009jz,AguilarSaavedra:2011zy} , but this signal is absent if a second, degenerate $Z'$ is incorporated with couplings such that the two extra bosons furnish a complex (reducible) representation. (Equivalently, the complex $Z'$ boson can be assigned a ``flavour'' charge.) In contrast, robust constraints on models for the $t \bar t$ asymmetry can be extracted from the measurements in $t \bar t$ production itself.

In view of the many possible new-physics scenarios and of the new measurements expected soon, it is pertinent and timely to find answers to the following questions:
\begin{enumerate}
\item[(a)] How well do simple models fit the Tevatron measurements?
\item[(b)] What are the implications of the new LHC data on the parameter space of these models?
\item[(c)] What are their predictions for the charge asymmetries at LHC?
\end{enumerate}

We have partly addressed questions (b) and (c) in previous work~\cite{AguilarSaavedra:2011hz}, focusing on model discrimination by the simultaneous measurement of the FB asymmetry at Tevatron and the charge asymmetry at LHC. In this paper we continue that analysis in several directions. First, we study (a) by comparing how different models fit the inclusive FB asymmetry measurement as well as the ones for high mass and high rapidity. In relation to (b), we investigate the effect of a precise measurement of the $t \bar t$ tail, which would be dramatic for the $Z'$ and $W'$ models in case that the measurement agrees with the SM expectation within a $\pm 50\%$. With respect to (c), we explore the dependence of the predictions on the mass of the new particle, as well as different definitions of charge asymmetries at LHC. This study, clearly, requires a scan over all the allowed values of the couplings and masses for each model, which goes beyond usual analyses with a few selected benchmark points. It is also crucial to impose existing constraints from experimental data, in order to bound the range of variation for the predictions.

\section{Simple models for the top asymmetry}
 
The new particles that contribute at tree level to $t\bar{t}$ production can be exchanged either in the $s$, $t$ or $u$ channels, depending on their couplings (which are conditioned by the gauge representation they belong to). The corresponding forms of the propagator have a significant impact on the invariant mass and rapidity distributions of the asymmetries and cross section. This can be used to classify the models in three groups:
\begin{itemize}
\item $s$-channel: For a heavy particle, both the cross section and asymmetries increase sharply with the invariant mass, more so in the region when the terms quadratic in the new physics are important. The effect becomes more acute as $m_{t\bar{t}}$ approaches the mass of the new particle.
\item $t$-channel: The propagator prefers forward top quarks, and increases the FB and charge asymmetries at high invariant masses and at large rapidities.
\item $u$-channel: The propagator prefers to send top quarks  {\em backwards}. Hence, it favours a negative asymmetry. Of course, this effect is counteracted in the proposed models by the numerator of the amplitude, so that a positive FB asymmetry is obtained at Tevatron. However, as the invariant mass increases, the rise of the asymmetries is damped by the influence of the propagator, and eventually the asymmetries become negative.
\end{itemize}
These behaviours of the propagators are milder when the mass of the exchanged particle increases, in relation to the Mandelstam variable in the denominator. In the heavy particle limit, the propagator becomes constant, and the effect can be described in all three cases by a four-fermion operator.

Another important distinction between the different models can be made according to whether the (positive) asymmetry excess is generated at linear or quadratic order in the new physics. It has been remarked in the literature that the agreement of Tevatron measurements of the total cross section with the SM prediction indicates that the new physics that enhances the FB asymmetry must interfere with the SM amplitudes~\cite{Grinstein:2011yv}. This condition is not very restrictive, however, as all the fields that can be exchanged at tree level in $q\bar{q} \to t\bar{t}$ do interfere with the SM, unless some couplings are set to zero. It is maybe more useful to distinguish two scenarios, depending on whether the terms that are quadratic in the new-physics amplitude are essential or not to generate the asymmetry: 
\begin{itemize}
\item {\em `Linear' new physics.} For sufficiently small coupling/mass ratio, the impact of the terms quadratic in the new physics amplitude is small, and interference with the SM dominates. In this class of models, the interference contribution to the cross section must cancel or be small. Among simple models, the former is only possible for a colour-octet vector boson with axial couplings either to the top or the light quarks\footnote{In order to have a vanishing interference with the tree-level QCD amplitude it is sufficient that one of the couplings is axial; however, in order to have a non-zero asymmetry both couplings must have an axial component. The asymmetry generated is maximal when both the top and light quark couplings are purely axial.}; the latter happens for a colour-sextet scalar. These models have the feature that the coupling can be smoothly varied between zero and some (mass-dependent) maximum while keeping agreement with the total $t \bar t$ cross section, for example.
\item {\em `Quadratic' new physics.} These require large coupling/mass ratio, in order to make quadratic and interference terms comparable. Then, the contributions of the latter to the cross section must be negative, such as to cancel the quadratic contribution. The problem with this scenario is that the cancellation imposed at a given energy is not guaranteed at different energies. Hence, an agreement with the Tevatron cross section typically leads to a significant excess in the LHC $t\bar{t}$ tail. They also have the feature that the allowed region for the couplings that generate a positive asymmetry is not connected with the SM point of vanishing couplings; therefore, these models can easily be disfavoured by the upcoming precise LHC measurements.
\end{itemize}

In this paper we study six models that are representative of these alternatives. They cover the different tree-level exchanges, and the scenarios with linear and quadratic new physics.  Furthermore, the $d\bar{d}$ and $u\bar{u}$ initial states are both represented. Our selection comprises the popular models mentioned in the introduction, plus a flavour-violating scalar isodoublet~\cite{Shu:2009xf,Nelson:2011us,Zhu:2011ww,Cao:2011yt,Babu:2011yw}. The possible quantum numbers and relevant interactions of the extra fields are collected in Table~\ref{tab:lagr}. We follow the notation in~\cite{AguilarSaavedra:2011vw,delAguila:2010mx}  for the different irreducible representations of the complete gauge group.\footnote{In particular, the symbol $\bmu$ to denotes an extra vector boson which is neutral under the full SM gauge group, whereas the name $Z^\prime$ refers to any colour-singlet, electrically neutral vector boson, which could belong to different isospin multiplets. Similarly,  the charge~$\pm 1$ vector bosons $W'$ can be isosinglets $\bmu^1$ or components of isotriplets $\mathcal{W}_\mu$. In this paper we only consider isosinglet $Z'$ and $W'$ bosons, but we still make these distinctions because the couplings of the new fields keep track of their complete quantum numbers.}.
\begin{table}[htb]
\begin{center}
\begin{tabular}{ccl}
Label & Rep. & \multicolumn{1}{c}{Interaction Lagrangian} \\
\hline
$\bmu$ & $(1,1)_0$ 
  & $-\left( g_{ij}^q \bar q_{Li} \gM q_{Lj} 
  + g_{ij}^u \bar u_{Ri} \gM u_{Rj} \right.$ \\
  & & $\left. + g_{ij}^d \bar d_{Ri} \gM d_{Rj} \right) \bmu $ \\[1mm]
$\bmu^1$ & $(1,1)_1$ 
  & $- g_{ij} \bar d_{Ri} \gM u_{Rj} \, \bmu^{1\dagger} + \text{h.c.}$ \\[1mm]
$\gmu$ & $(8,1)_0$
  & $- \left( g_{ij}^q \bar q_{Li} \gM \frac{\la}{2} q_{Lj} 
  + g_{ij}^u \bar u_{Ri} \gM \frac{\la}{2} u_{Rj} \right.$ \\
  & & $\left. + g_{ij}^d \bar d_{Ri} \gM \frac{\la}{2} d_{Rj} \right) \mathcal{G}_\mu^a$ \\[1mm]
$\phi$ & $(1,2)_{-\frac{1}{2}}$
  & $- g_{ij}^u \bar q_{Li} u_{Rj} \, \phi - g_{ij}^d \bar q_{Li} d_{Rj} \, \tilde \phi  + \text{h.c.}$ \\[1mm]
$\omf$ & $(3,1)_{-\frac{4}{3}}$
  & $- g_{ij} \varepsilon_{abc} \bar u_{Rib} u_{Rjc}^c \, \omega^{4a\dagger} + \text{h.c.}$ \\[1mm]
$\OMf$ & $(\bar 6,1)_{-\frac{4}{3}}$
  & $-g_{ij} \oh \left[ \bar u_{Ria} u_{Rjb}^c + 
  \bar u_{Rib} u_{Rja}^c \right] \Omega^{4ab\dagger} + \text{h.c.}$ \\
\end{tabular}
\caption{Selection of vector bosons and scalar representations mediating $q \bar q \to t \bar t$ and their relevant interactions.\label{tab:lagr}}
\end{center}
\end{table}
More specifically, we consider the following models:
\begin{enumerate}
\item $Z'$ boson: A neutral (colour- and isospin-singlet) vector boson $\bmu$, exchanged in the $t$ channel in the process $u \bar u \to t \bar t$. We take the $Z'tu$ couplings to be right-handed, only $g_{13}^u \neq 0$, as preferred by bounds from $B$ physics. Our results are independent of this choice, anyway. For a real $Z'$ field, the contribution to the FB and charge asymmetries is strongly constrained by the nonobservation of like-sign top pair production~\cite{AguilarSaavedra:2011zy,CDFtt,Collaboration:2011dk} but, as we have mentioned, these bounds can be evaded (see~\cite{Jung:2009jz}).

\item $W'$ boson: A charged (colour- and isospin singlet) vector  $\bmu^1$, with right-handed couplings $g_{13}$, exchanged in the $t$ channel in $d \bar d \to t \bar t$. Charged bosons with left-handed couplings could appear as components of $\text{SU}(2)_L$ triplets, but this possibility is again disfavoured by $B$ physics constraints.

\item Axigluon: A neutral colour-octet vector $\gmu$ with axial couplings $g_{ii}^q = - g_{ii}^u = - g_{ii}^d$, exchanged in the $s$ channel in $q \bar q \to t \bar t$. We consider this new particle to be heavy enough not to be produced on shell, to avoid strong bounds from the $m_{t\bar{t}}$ profile. A large width has been invoked in Ref.~\cite{Barcelo:2011vk} (see also~\cite{Tavares:2011zg,Alvarez:2011hi,AguilarSaavedra:2011ci}) to hide light axigluon resonances, but we do not include this case in our general analysis of simple models because the results are strongly sensitive to the details of the model. For instance, in the stealth gluon model proposed in~\cite{Barcelo:2011vk}, the large width arises from the decay to a light vector-like quark. Nevetheless, we will make some remarks about light axigluons later on, and show some results for a specific model.

\item Scalar doublet: A colour-singlet Higgs-like isodoublet $\phi$,  which contains neutral and charged scalars, coupling the top quark to the first generation and exchanged in the $t$ channel. The presence of both couplings $g_{13}^u$ and $g_{31}^u$ leads to like-sign top production~\cite{AguilarSaavedra:2011zy} so one of these couplings must be negligible to avoid these constraints. We consider $g_{31}^u$ non-zero, which is preferred by low-energy constraints~\cite{Blum:2011fa}. The alternative with $g_{13}^u$ non-zero produces quite similar results. We neglect the effect (possibly relevant for small masses) of possible splittings of the two components induced by electroweak breaking, as they are very model-dependent.

\item Colour-triplet scalar: A charge 4/3 colour-triplet $\omf$ with flavour-violating $tu$ couplings $g_{13}$, necessarily right-handed, exchanged in the $u$ channel in $u \bar u \to t \bar t$. Notice that the antisymmetry in colour indices implies that diagonal couplings to $uu$, $tt$ identically vanish. 

\item Colour-sextet scalar: A charge 4/3 colour-sextet $\OMf$, also with right-handed flavour-violating $tu$ couplings $g_{13}$, and exchanged in the $u$ channel. In contrast with $\omf$, for the sextet there may be diagonal $uu$, $tt$ couplings, albeit not related to the flavour-violating ones. They can potentially give rise to large (unobserved) $tt$ signals unless suppressed by some flavour symmetry~\cite{Grinstein:2011yv,Ligeti:2011vt}.
\end{enumerate}

\section{Asymmetries at Tevatron}

The CDF Collaboration has reported in Ref.~\cite{Aaltonen:2011kc} a measurement of the inclusive FB asymmetry as well as for high $t \bar t$ invariant masses and large $\Delta y$ rapidity differences,
\begin{align}
& A_\text{FB}^\text{exp} =  0.158 \pm 0.075 && \text{(inclusive)} \,, \notag \\
& A_\text{FB}^\text{exp} =  0.475 \pm 0.114 && (m_{t\bar t} > 450~\text{GeV}) \,, \notag \\
& A_\text{FB}^\text{exp} =  0.611 \pm 0.256 && (|\Delta y| > 1) \,,
\label{ec:afbX}
\end{align}
all at the parton level. To study how the different models can reproduce these values, we scan over the masses from 100 GeV to 10 TeV and arbitrary couplings, and select the values that satisfy two constraints. (Extending the range to lower masses does not change the results for $Z'$, $W'$, $\omf$ and $\OMf$, nor affects the essential conclusions for a scalar doublet.) First, we require agreement with the total $t \bar t$ cross section measured at Tevatron, $\sigma = 7.50 \pm 0.48$ pb~\cite{CDFtt}. 
There are different calculations of the SM cross section. While most results lie pretty close to the measurement, for example $\sigma = 7.46^{+0.66}_{-0.80}$ pb~\cite{Langenfeld:2009wd}, significantly smaller values, $\sigma = 6.30 \pm 0.19^{+0.31}_{-0.23}$, have also been found~\cite{Ahrens:2010zv}. Taking into account the uncertainties in the theoretical predictions and in the experimental measurement, we demand that new physics contributions to $t \bar t$ production should lie inside the interval $[-0.8,1.7]$ pb, as in Ref.~\cite{AguilarSaavedra:2011hz}. 
Second, we restrict the impact on the $t \bar t$ tail at LHC. There are not yet specific measurements of the high-mass $t \bar t$ cross section, but the observed distributions~\cite{CMStail} agree well with the SM prediction. Then, in our analysis we require that the cross section for $m_{t \bar t} > 1$ TeV is at most three times its SM value. In the next section we study the effect of tightening this constraint. We do not attempt to reproduce the the Tevatron $t \bar t$ invariant mass distribution, but we note that for most of the parameter space left by the previous constraints the agreement is good~\cite{Jung:2009jz,Ligeti:2011vt,Grinstein:2011yv}, partly because of a smaller efficiency for the new physics contributions~\cite{Gresham:2011pa}. All cross sections are calculated with the tree-level generator {\tt Protos}~\cite{AguilarSaavedra:2008gt}.

\begin{figure}[p]
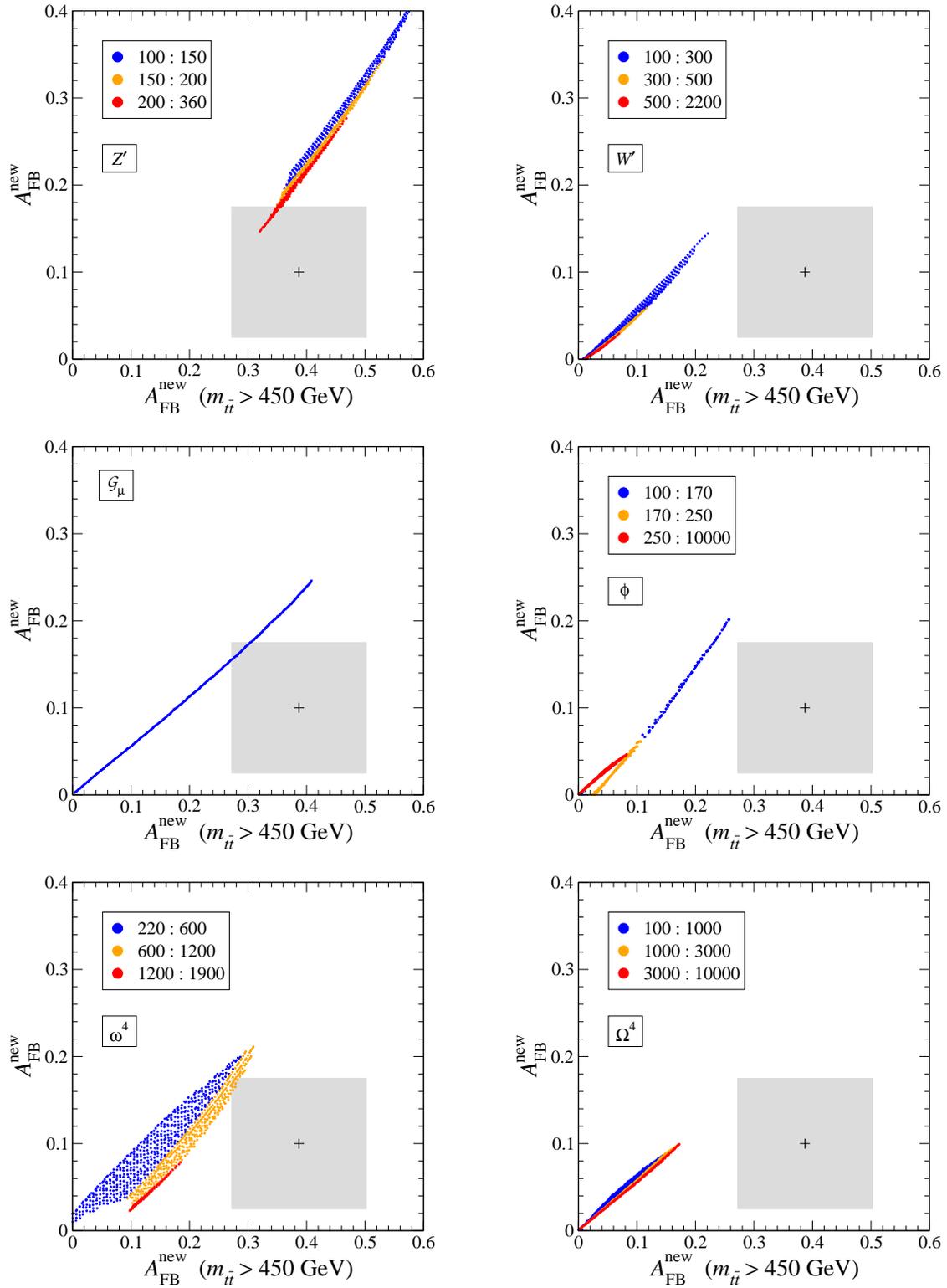

\begin{center}
\begin{tabular}{ccc}
\epsfig{file=Figs/T1T0-Zp.eps,width=6.7cm,clip=} & \quad\quad &
\epsfig{file=Figs/T1T0-Wp.eps,width=6.7cm,clip=} \\[2mm]
\epsfig{file=Figs/T1T0-Ax.eps,width=6.7cm,clip=} & &
\epsfig{file=Figs/T1T0-ph.eps,width=6.7cm,clip=} \\[2mm]
\epsfig{file=Figs/T1T0-om.eps,width=6.7cm,clip=} & &
\epsfig{file=Figs/T1T0-Om.eps,width=6.7cm,clip=}
\end{tabular}
\caption{Allowed regions for the new physics contributions to the high-mass and inclusive asymmetries at Tevatron. The mass ranges in the legends are in GeV.}
\label{fig:T1T0}
\end{center}
\end{figure}

\begin{figure}[p]
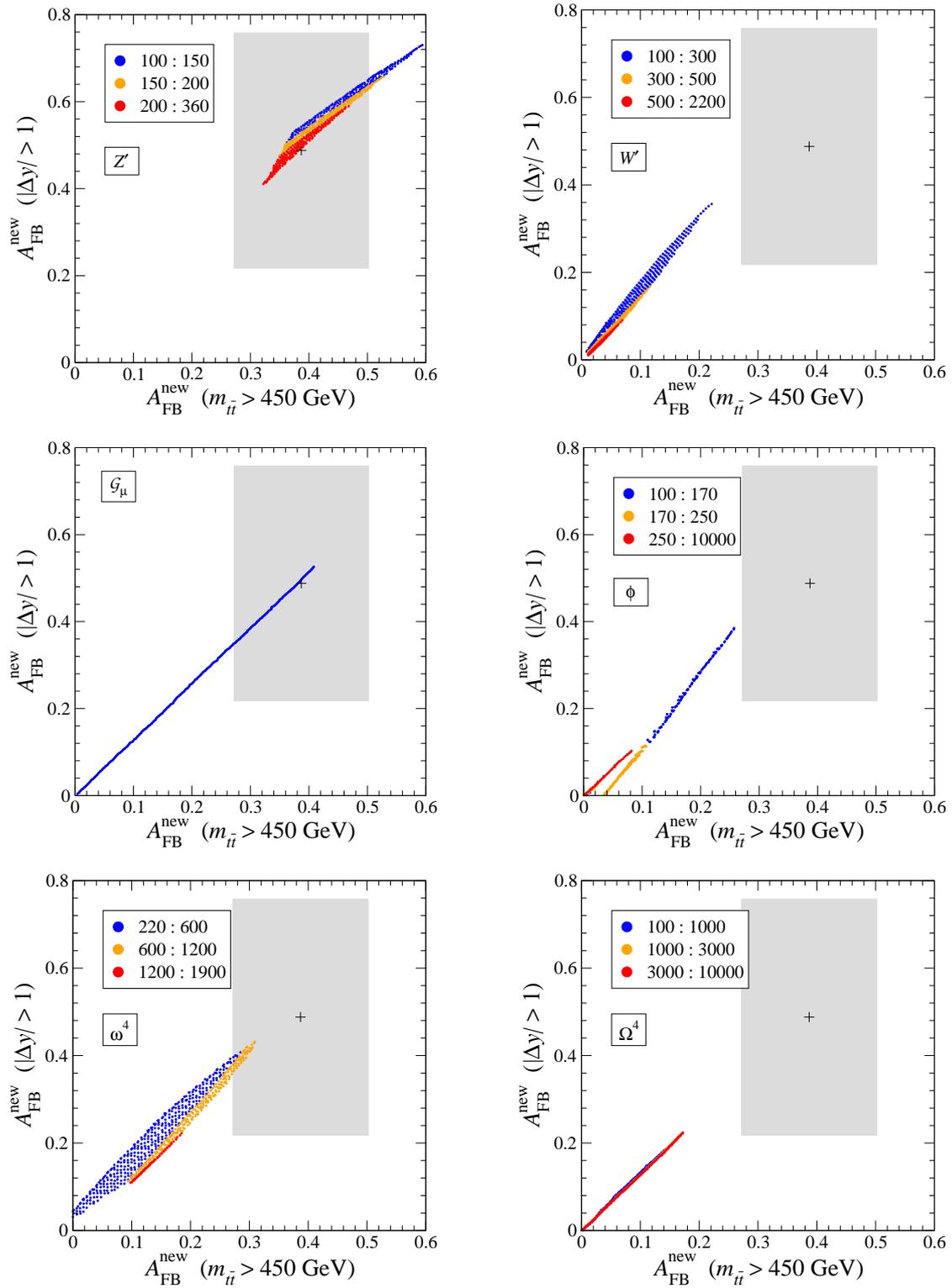

\begin{center}
\begin{tabular}{ccc}
\epsfig{file=Figs/T1Ty-Zp.eps,width=6.7cm,clip=} & \quad\quad &
\epsfig{file=Figs/T1Ty-Wp.eps,width=6.7cm,clip=} \\[2mm]
\epsfig{file=Figs/T1Ty-Ax.eps,width=6.7cm,clip=} & &
\epsfig{file=Figs/T1Ty-ph.eps,width=6.7cm,clip=} \\[2mm]
\epsfig{file=Figs/T1Ty-om.eps,width=6.7cm,clip=} & &
\epsfig{file=Figs/T1Ty-Om.eps,width=6.7cm,clip=}
\end{tabular}
\caption{Allowed regions for the new physics contributions to the high-mass and high $\Delta y$ asymmetries at Tevatron. The mass ranges in the legends are in GeV.}
\label{fig:T1Ty}
\end{center}
\end{figure}

For clarity, instead of giving our predictions in terms of the total asymmetries, we consider the contributions $\afb^\text{new}$ from new physics beyond the SM. The total asymmetries are obtained with good approximation by summing to these values the SM ones~\cite{Campbell:1999ah},
$\afb^\text{SM} = 0.058 \pm 0.009$ (inclusive)
$\afb^\text{SM} = 0.088 \pm 0.013$ ($m_{t\bar t} > 450~\text{GeV}$)
$\afb^\text{SM} = 0.123 \pm 0.018$ ($|\Delta y| > 1$).
We present in Fig.~\ref{fig:T1T0} the possible predictions of the six models in the plane of high-mass and inclusive FB asymmetries, once the constraints have been imposed.  Similarly, we plot in Fig.~\ref{fig:T1Ty} the allowed regions for the contributions of these models to the asymmetry at high $m_{t \bar t}$ versus the one at high rapidity difference. 

The results obtained for each model deserve a separate discussion.
For a $Z'$ boson, the plot shows that the allowed parameter space yielding positive contributions to the asymmetry is strongly constrained by the agreement with the Tevatron cross section and LHC tail~\cite{AguilarSaavedra:2011hz}.  In fact, because the interference with the SM is negative, the exchange of a $Z'$ decreases the FB asymmetry at first order~\cite{AguilarSaavedra:2011vw} so that a large coupling is required for a positive $\afb^\text{new}$, such that the quadratic terms dominate. But there is just a small window for large couplings that keep the agreement with the total cross section measured at Tevatron. The additional requirement of a small tail at LHC further reduces the parameter space, and eventually a $Z'$ mass $M \leq 360$ GeV is required. For masses below this limit, a $Z'$ boson can reproduce well the CDF values of the asymmetries, as shown in Figs.~\ref{fig:T1T0} and \ref{fig:T1Ty}. A striking consequence of the analysis, manifest in the plot, is that there are {\em minimum} values of the FB asymmetries for this model~\cite{AguilarSaavedra:2011hz}. The reason is the need of a large coupling to achieve the cancellation between interference and quadratic terms. Therefore, if the high-mass asymmetry turns out to be significantly smaller than its present central value, this model will be disfavoured. 

The $W'$ boson has negative interference as well, and again requires a large coupling to yield a positive $\afb^\text{new}$. The mass range is less constrained in this case: $M_{W'} \leq 2.2$ TeV. The main difference with a $t$-channel $Z'$ boson
is that the relevant process here, $d \bar d \to t \bar t$, is relatively more important at LHC than at Tevatron, so that the constraints from the LHC tail have a larger impact. As a result, the possible contributions to the asymmetry are smaller than for a $Z'$, especially at high invariant masses $m_{t \bar t} > 450$ GeV. The mass range where the high-mass asymmetry measurement can be accommodated within two standard deviations is $M_{W'} \lesssim 300$ GeV. We also point out that, in contrast with a $Z'$ boson, small positive FB asymmetries can be obtained while keeping agreement with the Tevatron cross section.

A heavy axigluon can fit well the three asymmetry measurements, as shown in Figs.~\ref{fig:T1T0} and \ref{fig:T1Ty}. 
It is important that this is a model with linear new physics, so the good fit does not need large cancellations between linear and quadratic terms, as it happens for a $Z'$ or $W'$. In this sense the model is more natural and robust. As long as the axigluon is heavy enough, it does not produce too large an LHC tail~\cite{AguilarSaavedra:2011vw}. The generated asymmetry is proportional to the product of the couplings to the light and top quark; since the former is constrained to be small by dijet cross section measurements, the latter must be somewhat large, though still perturbative, to reproduce the central value of the CDF measurement. Note also that, if future $t \bar t$ resonance searches do not find a positive signal and push further up the mass limit for axigluons, the required coupling to the top quark may be too large.

Scalar isodoublets have seldom been discussed in the literature in the context of the FB asymmetry, but it turns out that they give a reasonable fit and can pass the LHC constraints better than other more popular models. In this case, the interference is positive and enhances the asymmetry, while quadratic terms dilute it. For large masses and only one coupling ($g_{13}^u$ or $g_{31}^u$), as required by limits on like-sign $tt$, the asymmetries are small \cite{AguilarSaavedra:2011zy}. However, for small masses, up to $M \sim 250$ GeV, there is a window of large couplings giving a large asymmetry, a small LHC tail and agreement with the Tevatron cross section (with cancellation of linear and quadratic terms). In particular, for masses up to 170 GeV, this model can accommodate the inclusive asymmetry and values slightly smaller for the other two measurements in Eqs.~(\ref{ec:afbX}). For such masses, however, one has to worry about flavour-changing decays of the top quark~\cite{AguilarSaavedra:2000aj}, so building a realistic model may not be easy. Besides, for light scalars small couplings are also possible. The allowed regions for the asymmetries approximately coincide with the region for $M > 250$ GeV (shown in red in the plots).

The scalar triplet $\omf$ contributes to $u \bar u \to t \bar t$, with negative interference with the SM~\cite{Arhrib:2009hu,Ligeti:2011vt}. Once again, it requires large couplings and a cancellation to explain the measured asymmetries. In contrast with a $Z'$ boson, which also contributes to this process, the new particle is exchanged in the $u$ channel. For very light scalars, the preference of the propagator for backward top quarks is enhanced, which forbids positive contributions to the asymmetry, while for masses $M \geq 220$ GeV the numerator compensates this effect. For masses above $M \sim 1$ TeV the size of the contributions $\afb^\text{new}$ is very constrained by the LHC tail, and the model cannot accommodate the measured asymmetries.\footnote{
For the colour-triplet and sextet scalars the tail enhancement found in Ref.~\cite{AguilarSaavedra:2011vw} is reached only for large masses; light scalars do not produce such an effect.} Note also that for heavier masses there is a minimum positive FB asymmetry when agreement with the measured cross section is required, as for a $Z'$ boson.

Finally, for the scalar sextet $\OMf$ the interference is positive, and a positive asymmetry can naturally be produced at linear order. Still, the generated asymmetries are too small to explain the central values of the high-mass and high-$\Delta y$ measurements in Eqs.~(\ref{ec:afbX}).

\section{Consequences of a precise measurement of the LHC tail}

\begin{figure}[p]
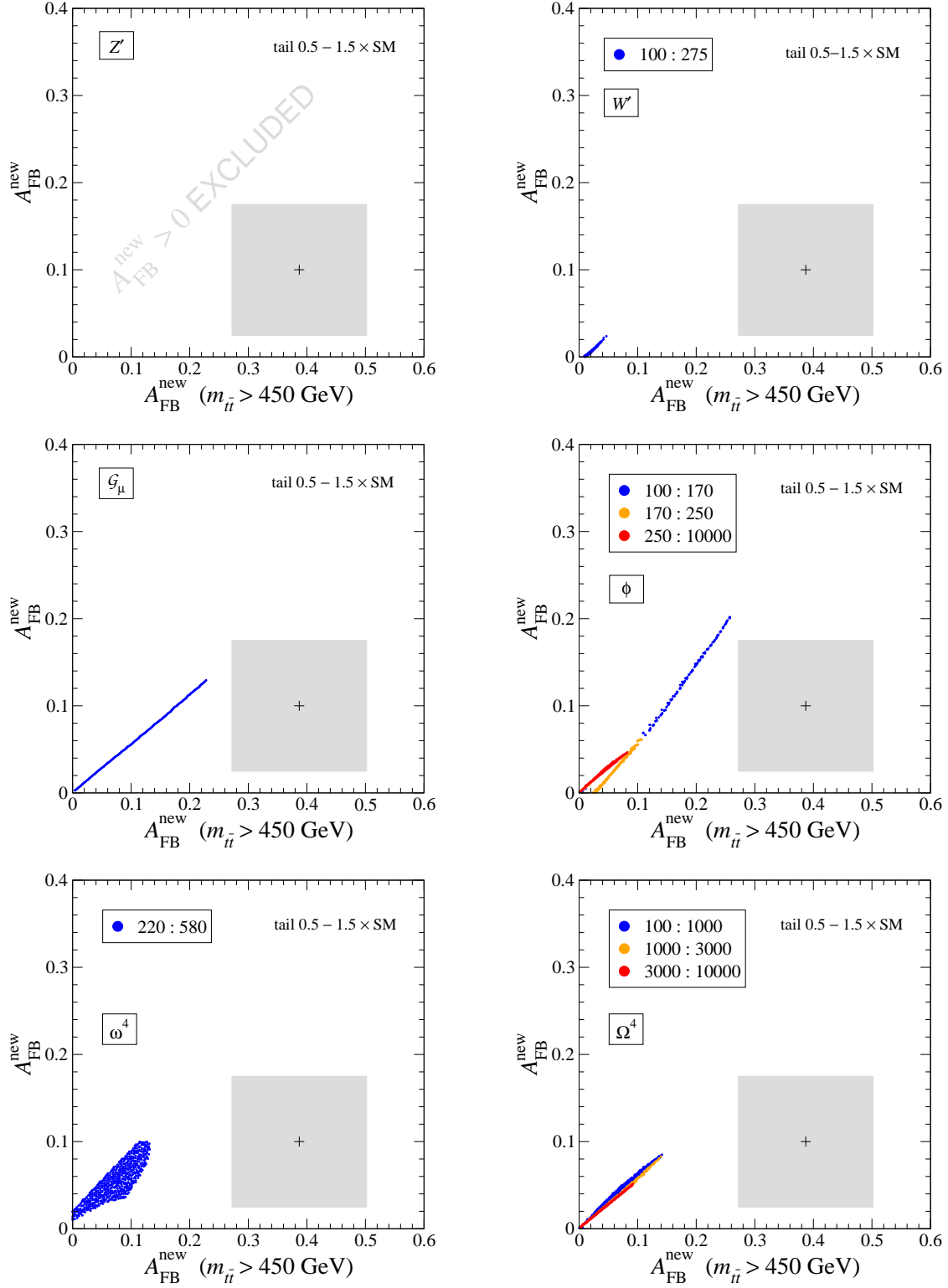

\begin{center}
\begin{tabular}{ccc}
\epsfig{file=Figs/T1T0i-Zp.eps,width=6.7cm,clip=} & \quad\quad &
\epsfig{file=Figs/T1T0i-Wp.eps,width=6.7cm,clip=} \\[2mm]
\epsfig{file=Figs/T1T0i-Ax.eps,width=6.7cm,clip=} & &
\epsfig{file=Figs/T1T0i-ph.eps,width=6.7cm,clip=} \\[2mm]
\epsfig{file=Figs/T1T0i-om.eps,width=6.7cm,clip=} & &
\epsfig{file=Figs/T1T0i-Om.eps,width=6.7cm,clip=}
\end{tabular}
\caption{Allowed regions for the new physics contributions to the high-mass and inclusive asymmetries at Tevatron, with an improved measurement of the LHC tail. The mass ranges in the legends are in GeV.}
\label{fig:T1T0i}
\end{center}
\end{figure}

Once the high-mass $t \bar t$ tail is precisely measured at LHC, the picture described in the previous section may change significantly. For illustration, we show in Fig.~\ref{fig:T1T0i} the allowed regions for the high-mass and inclusive asymmetries, if the cross section $\sigma(m_{t \bar t} > 1~\text{TeV})$ is required to be within $\pm 50\%$ of its SM value. (The lower bound only affects the scalar $\OMf$, since the rest of models always give a tail enhancement.)
As it is apparent, the consequences of the increased precision are dramatic. If agreement with the SM is found in the tail, positive contributions to the asymmetry are completely excluded for a $Z'$ boson; actually, they are even excluded if the tail is, at most, twice the SM prediction. For a $W'$ positive values of $\afb^\text{new}$ are still allowed, but must be tiny. The colour-triplet and sextet scalars would not be able to produce a high-mass FB asymmetry within $2\sigma$ of the CDF measurement. 

The only simple models accommodating this measurement to some extent would be either a heavy axiguon or a very light scalar doublet, which also predicts a small tail at LHC. Note, however, that for axigluons with masses around the TeV, the cross section enhancement is very large, as they are produced in the $s$ channel, and these new particles should be visible or excluded soon. On the other hand, a concealed light axigluon would not give any enhancement of the tail~\cite{Barcelo:2011vk,AguilarSaavedra:2011ci}.

\section{Predictions for charge asymmetries at LHC}

The measurement of a charge asymmetry at LHC would be an independent, crucial confirmation of the excesses found at Tevatron. Furthermore, when precisely measured, LHC charge asymmetries will give important information about the possible new physics models. As discussed in detail in Ref.~\cite{AguilarSaavedra:2011hz}, the combination of the measurements of the charge asymmetry at LHC and the FB asymmetry at Tevatron can be used to discriminate among the different models. Here, we study other aspects of the predictions of simple models for the charge asymmetries.

\begin{figure}[p]
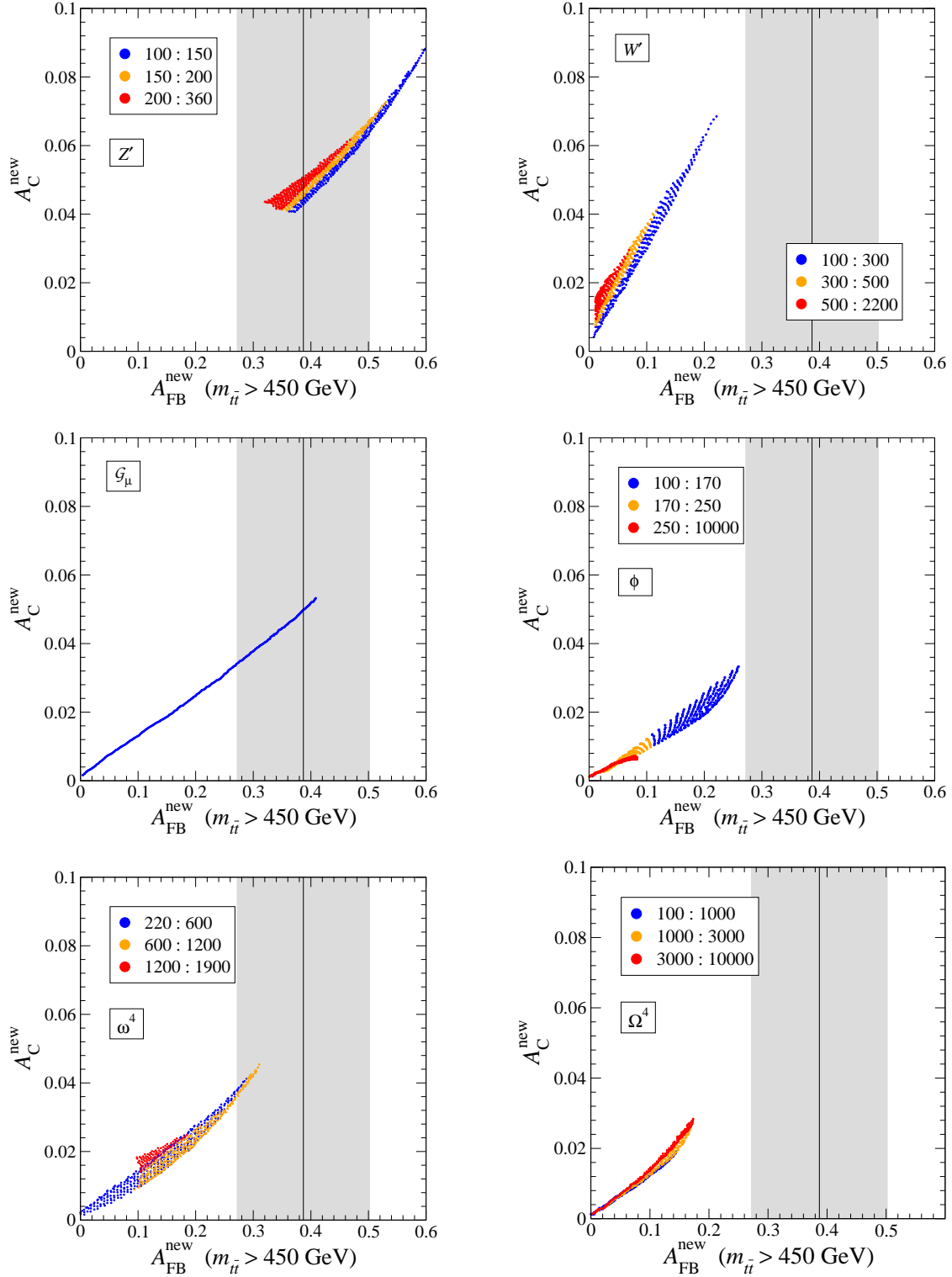

\begin{center}
\begin{tabular}{ccc}
\epsfig{file=Figs/T1L0-Zp.eps,width=6.7cm,clip=} & \quad\quad &
\epsfig{file=Figs/T1L0-Wp.eps,width=6.7cm,clip=} \\[2mm]
\epsfig{file=Figs/T1L0-Ax.eps,width=6.7cm,clip=} & &
\epsfig{file=Figs/T1L0-ph.eps,width=6.7cm,clip=} \\[2mm]
\epsfig{file=Figs/T1L0-om.eps,width=6.7cm,clip=} & &
\epsfig{file=Figs/T1L0-Om.eps,width=6.7cm,clip=}
\end{tabular}
\caption{Allowed regions for the new physics contributions to the high-mass FB asymmetry at Tevatron and the inclusive charge asymmetry at LHC. The mass ranges in the legends are in GeV.}
\label{fig:T1L0}
\end{center}
\end{figure}

\begin{figure}[p]
\begin{center}
\begin{tabular}{ccc}
\epsfig{file=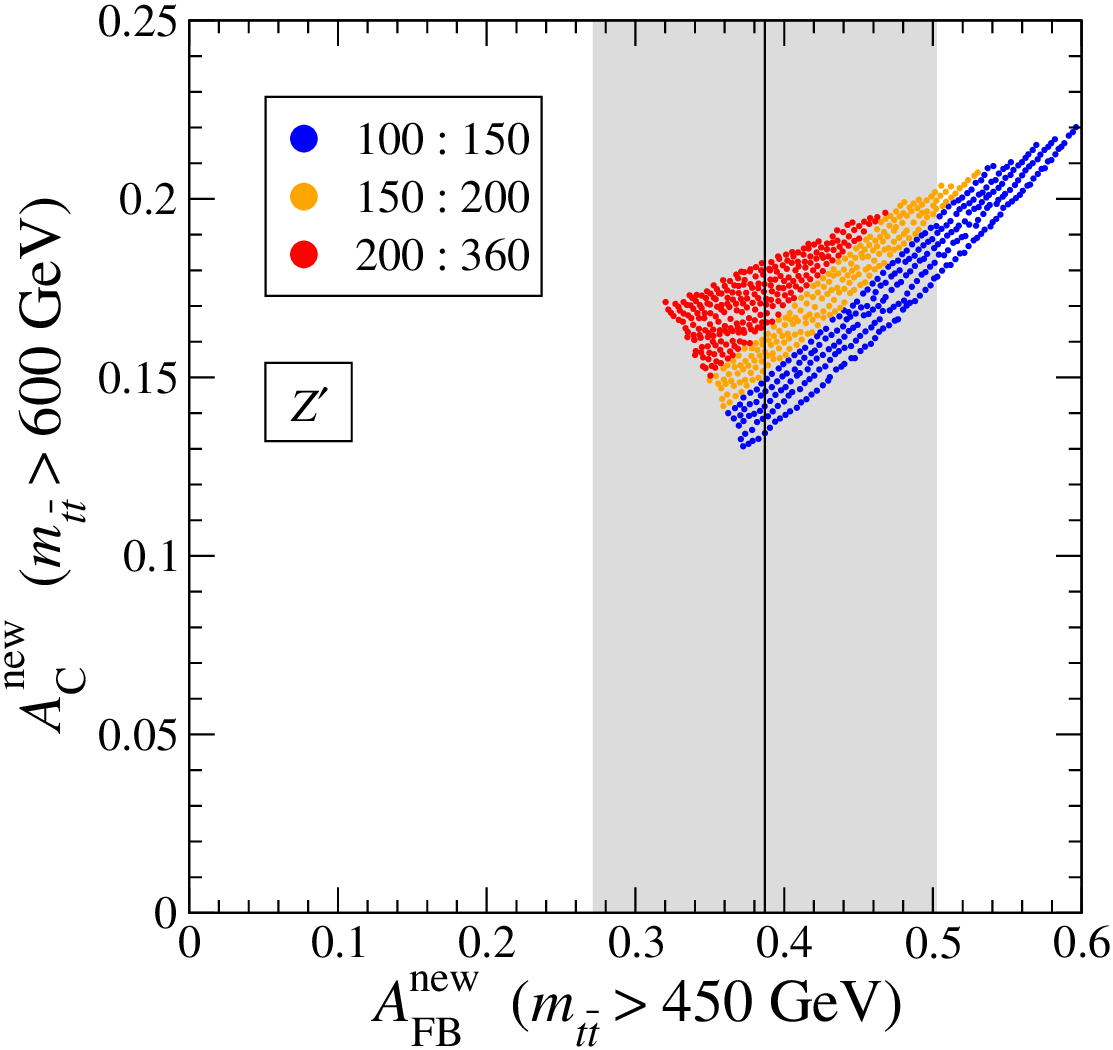,width=6.7cm,clip=} & \quad\quad &
\epsfig{file=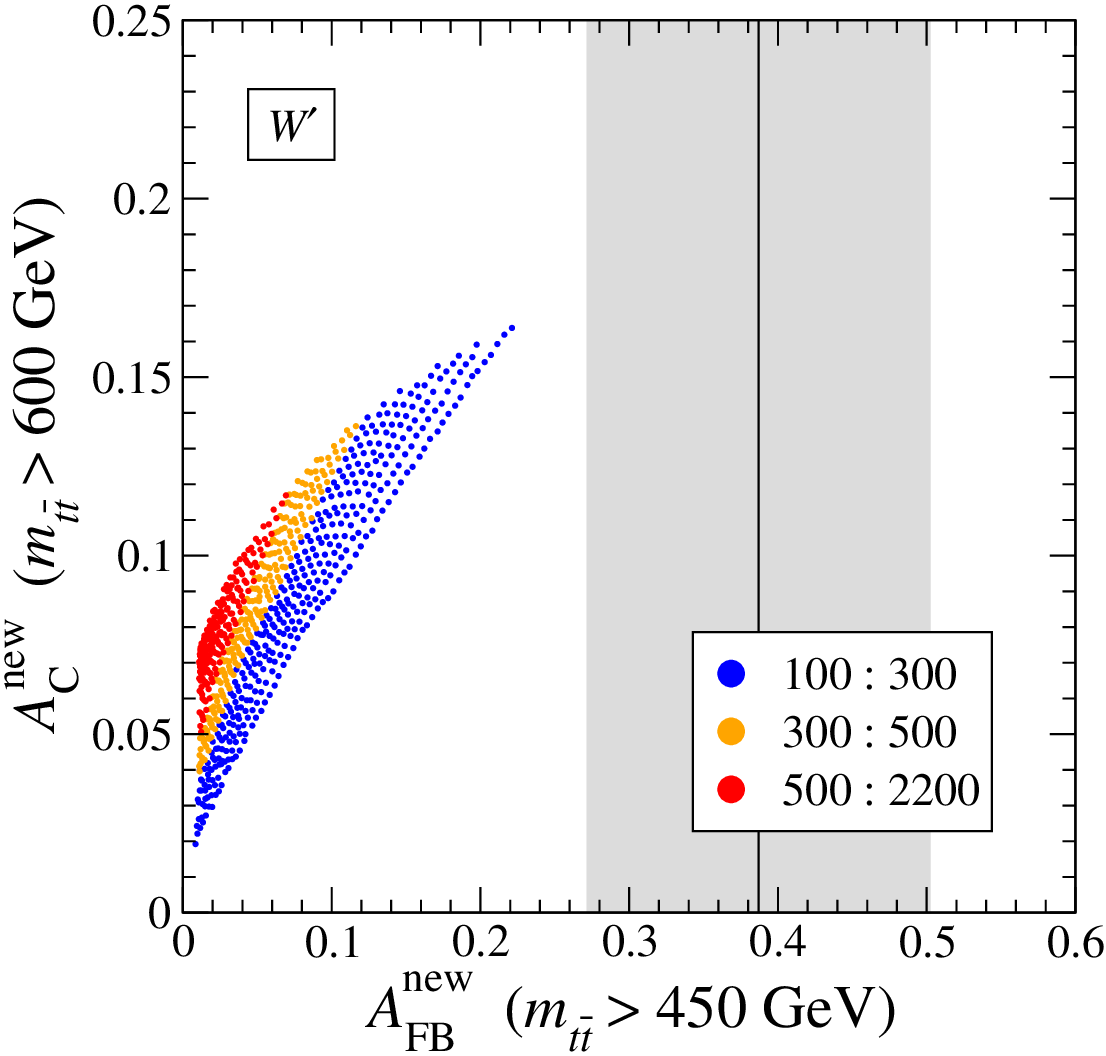,width=6.7cm,clip=} \\[2mm]
\epsfig{file=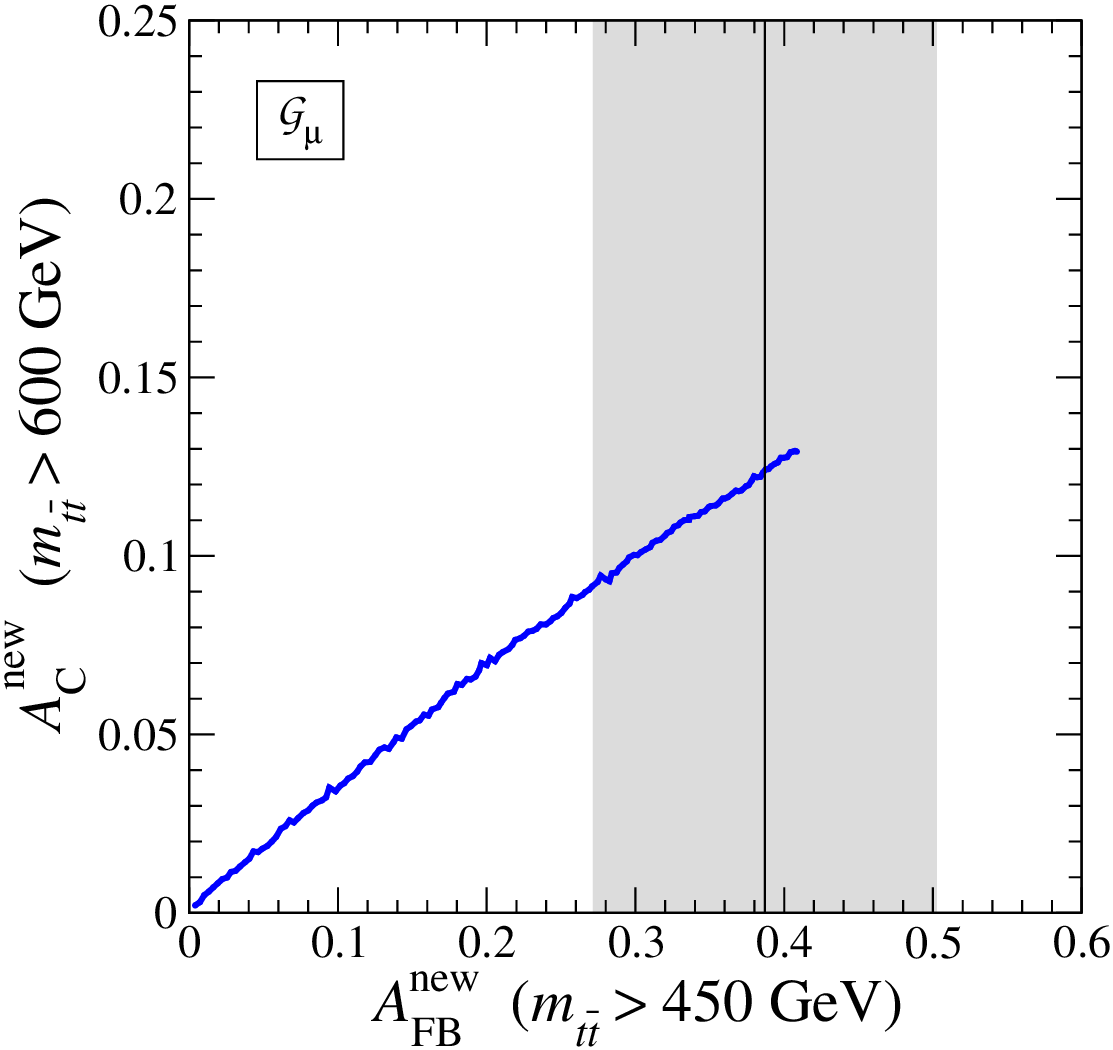,width=6.7cm,clip=} & &
\epsfig{file=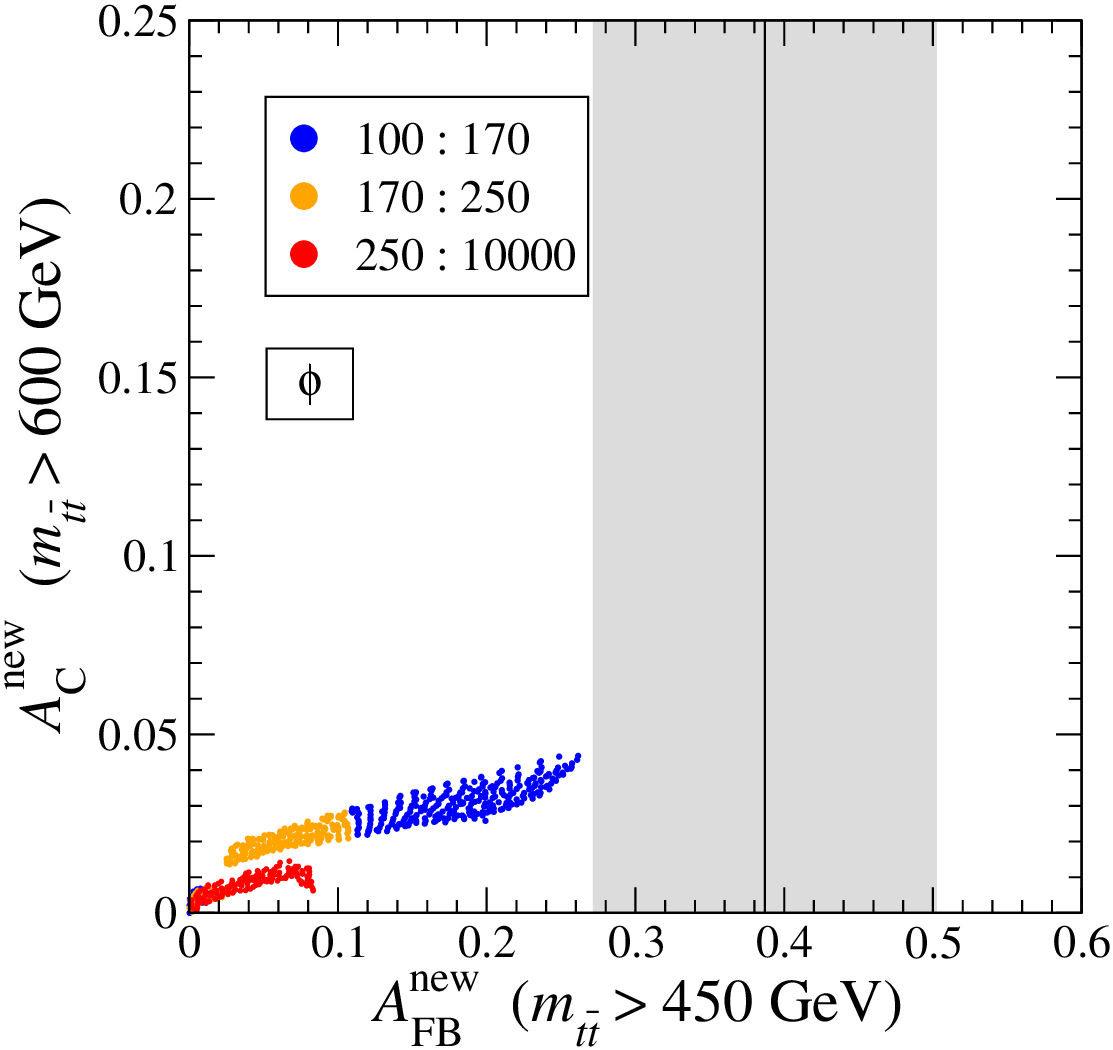,width=6.7cm,clip=} \\[2mm]
\epsfig{file=Figs/T1L1-om.eps,width=6.7cm,clip=} & &
\epsfig{file=Figs/T1L1-Om.eps,width=6.7cm,clip=}
\end{tabular}
\caption{Allowed regions for the new physics contributions to the high-mass FB asymmetry at Tevatron and the high-mass charge asymmetry at LHC. The mass ranges in the legends are in GeV.}
\label{fig:T1L1}
\end{center}
\end{figure}

The CMS Collaboration defines the charge asymmetry in $t\bar{t}$ production as~\cite{afbCMS}
\begin{equation}
\ac = \frac{N(\Delta > 0) - N(\Delta < 0)}{N(\Delta > 0) + N(\Delta < 0)} \,,
\label{ec:ACMS}
\end{equation}
where $\Delta = |\eta_t|- |\eta_{\bar t}|$, with $\eta_p$ the pseudo-rapidity of the particle $p$ in the laboratory frame, and $N(c)$ the number of $t \bar{t}$ events satisfying the condition $c$. Using $\Delta = |y_t|- |y_{\bar t}|$, with $y_p$ the rapidity of $p$, leads to the same value of $\ac$, and this is also the case for a FB asymmetry defined by taking the forward direction as the one of the longitudinal boost of the $t \bar t$ system~\cite{Krohn:2011tw}.
We present in Fig.~\ref{fig:T1L0} the predictions of our six simple models for this charge asymmetry.  In Fig.~\ref{fig:T1L1}, we show the predictions for the same charge asymmetry when restricted to events with invariant masses $m_{t \bar t} > 600$ GeV.


\begin{figure}[htb]
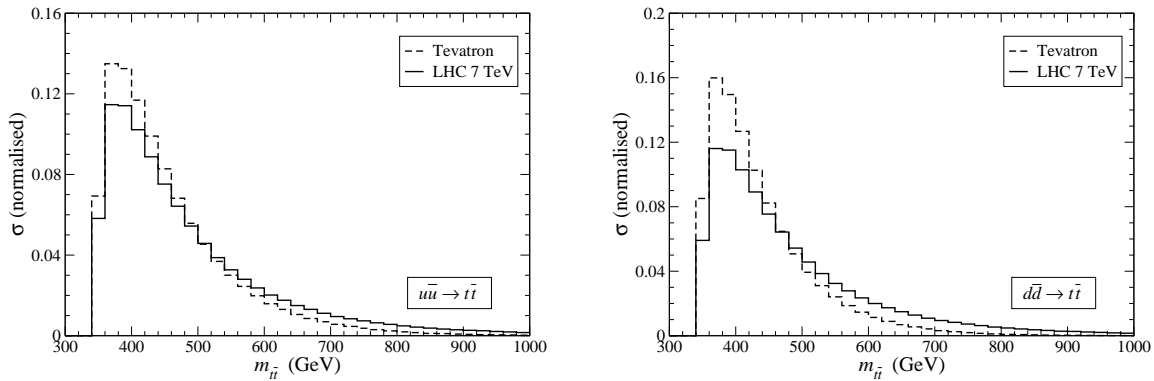

\begin{center}
\begin{tabular}{ccc}
\epsfig{file=Figs/mtt-uu.eps,width=7.1cm,clip=} & \quad&
\epsfig{file=Figs/mtt-dd.eps,width=7.1cm,clip=}
\end{tabular}
\caption{Normalised $t \bar t$ invariant mass distribution in $u \bar u$ and $d \bar d$ subprocesses at Tevatron and LHC.}
\label{fig:mtt}
\end{center}
\end{figure}

A first clear feature seen in the plots is that, for the simple models studied here, a positive inclusive FB asymmetry at Tevatron leads to a positive inclusive charge asymmetry at LHC. This is a natural consequence of the similarity between the invariant mass distributions at Tevatron and LHC for the $u \bar u$ and $d \bar d$ subprocesses, which are depicted in Fig.~\ref{fig:mtt}. 
The central value of the current measurement of $\ac$ is positive, but it still has a large uncertainty. It is consistent with a vanishing value and does not yet show a clear preference for either sign. It is therefore appropriate to anticipate the consequences of the upcoming charge asymmetry measurements by CMS and ATLAS for different outcomes. To do this, we assume the same systematic uncertainty as the current CMS result in Eq.~(\ref{ec:acX}), and a statistical uncertainty scaled to 1 fb$^{-1}$, giving a total uncertainty of $\pm 0.036$. We can distinguish the following alternatives for the central value:
\begin{itemize}
\item A positive asymmetry, say $\ac = 0.05 \pm 0.036$, would strenghten the hints from the CDF and D0 Collaborations for the asymmetry, and would not place further constraints on our six models.
\item A small positive asymmetry, say $\ac = 0.02 \pm 0.036$, would put some tension in the $W'$ model when combined with the high-mass measurement in Eq.~(\ref{ec:afbX}). As it can be seen from Fig.~\ref{fig:T1L0}, a large FB asymmetry is associated with a large charge asymmetry in this model.
\item A small negative asymmetry, say $\ac = -0.02 \pm 0.036$, would clearly disfavour a $Z'$ boson, as it has a minimum value for the charge asymmetry, $\ac \geq 0.04$. For the rest of models it would lead to a tension with the high-mass CDF measurement.
\item A negative asymmetry, say $\ac = -0.05 \pm 0.036$, is unexpected and its understanding would be challenging.
\end{itemize}
At higher invariant masses the six models predict positive asymmetries, with an enhancement that depends on the specific model and also on the mass of the new particle. But having a positive high-mass asymmetry is not an absolute requirement. Indeed, some models predict that the asymmetry becomes negative at high masses~\cite{Barcelo:2011vk,AguilarSaavedra:2011ci}. For example, with a relatively light axigluon of mass $M=700$ GeV and couplings $g_{11}^q = -0.11$, $g_{33}^q = 5$, the following asymmetries are generated\footnote{For this mass and these couplings, the width is very large, $\Gamma \sim 0.66 M$, and its presence would be unnoticed in the $t \bar t$ invariant mass distributions both at Tevatron and LHC~\cite{AguilarSaavedra:2011ci}. Note that large top couplings (which may even be unperturbative) are not the only possibility to have a large width, which can also be achieved by opening new channels into additional extra particles~\cite{Barcelo:2011vk,Tavares:2011zg}. For definiteness, and since the results are independent of the particular way of enhancing the width, we have selected a benchmark with a large top coupling.}
\begin{align}
& \afb^\text{new} = 0.135 ~\text{(inclusive)} && \afb^\text{new} = 0.208 ~(m_{t \bar t} > 450~\text{GeV}) \,, \notag \\
& \ac^\text{new} = 0.018 ~\text{(inclusive)} && \ac^\text{new} = -0.005 ~(m_{t \bar t} > 600~\text{GeV}) \,.
\end{align}
These values are in sharp contrast with the predictions for the six models studied above. It should be clear that, as advocated in Ref.~\cite{AguilarSaavedra:2011hz}, it is of the utmost importance to study the $m_{t \bar t}$ dependence of the charge asymmetry at LHC. Thanks to the expected good statistics of $t\bar{t}$ pairs, this will be attainable in the near future.

Our plots also provide information about the mass dependence of the asymmetries in each model. As we see, the variation of the predictions with the mass of the new particle is much smaller than the differences between the predictions of the various models. This, of course, is a key ingredient for model discrimination~\cite{AguilarSaavedra:2011hz}. The mass dependence is enhanced when one considers the high-mass charge asymmetries, especially for the models involving $u$-channel exchange of $\omf$, $\OMf$. The reason for it has already been mentioned: the preference of the $u$-channel propagator for backward top quarks is more pronounced for higher invariant masses of the $t\bar{t}$ pair. Eventually, a precise measurement of the charge asymmetry at high $m_{t \bar t}$ could be use to extract information about the new particle mass.

Let us next explore other related definitions of charge asymmetries. Specifically, we consider the central~\cite{Ferrario:2008wm} and forward~\cite{Hewett:2011wz} charge asymmetries, which are defined, respectively, as
\begin{align}
& A_\text{cen} = \frac{N(|y_t| < y_C) - N(|y_{\bar{t}}| < y_C)}{N(|y_t| < y_C) + N_{\bar t}(|y_{\bar{t}}| < y_C)} \,, \notag \\[2mm]
& A_\text{fwd} = \frac{N(|y_t| > y_C) - N(|y_{\bar{t}}| > y_C)}{N(|y_t| > y_C) + N_{\bar t}(|y_{\bar{t}}| > y_C)} \,,
\label{ec:ACAF}
\end{align}
where the rapidity cut $y_C$ may be varied to optimise the sensitivity of these asymmetries. Considering statistical uncertainties only, the sensitivity is given by $A/\Delta A$, with
\begin{equation}
\Delta A = \frac{\sqrt{(\Delta N_t)^2 + (\Delta N_{\bar t})^2}}{N_t+N_{\bar t}} \,.
\end{equation}
The optimal value for $y_C$ depends on the new physics model considered. The values $y_C \simeq 0.7$ have been proposed for a heavy axigluon~\cite{Ferrario:2008wm}, while larger values $y_C \simeq 1.5$ are recommended for a light $Z'$~\cite{Hewett:2011wz}, as in this case the forward enhancement of the cross section is very pronounced. To illustrate the differences between the alternative definitions, we pick a benchmark point of each model (specified in Table~\ref{tab:bench}) and give in Table~\ref{tab:AC} the corresponding charge asymmetries, using the definition in Eq.~(\ref{ec:ACMS}) as well as those in Eqs.~(\ref{ec:ACAF}) with both $y_C = 1.5$ and $y_C = 0.7$. To allow for a fair comparison, we include in each case the figure of merit $A \times \sqrt \sigma \propto A/\Delta A$, where $\sigma$ corresponds to the cross section entering the denominator of the asymmetries.\footnote{Notice that for Gaussian statistics $\Delta N_{t,\bar t} = \sqrt{N_{t,\bar t}}$, so $\Delta A = (N_t + N_{\bar t})^{-1/2}$. These numbers of events are proportional to the cross sections $\sigma(|y_t| \lessgtr y_C)$, $\sigma(|y_{\bar t}| \lessgtr y_C)$. The actual sensitivities are obtained, for a given luminosity $\mathcal{L}$ and efficiency $\epsilon$, multiplying by $\sqrt{\epsilon \mathcal{L}}$.}
\begin{table}[htb]
\begin{center}
\begin{tabular}{cccc}
Model & Mass & Coupling & $\afb^\text{new}$ ($m_{t \bar t} > 450$ GeV) \\
$Z'$ & 150 GeV & $g_{13}^u = 0.55$ & 0.362 \\
$W'$ & 150 GeV & $g_{13} = 0.7$ & 0.175 \\
$\mathcal{G}_\mu$ & \multicolumn{2}{c}{$C/\Lambda^2 = 2$ TeV$^{-2}$} & 0.347 \\
$\phi$ & 120 GeV & $g_{31}^u = 1.4$ & 0.202 \\
$\omf$ & 400 GeV & $g_{13} = 1.5$ & 0.192 \\
$\OMf$ & 1 TeV & $g_{13} = 1.3$ & 0.148
\end{tabular}
\caption{Benchmark points selected to illustrate the sensitivities of the different charge asymmetries.}
\label{tab:bench}
\end{center}
\end{table}
\begin{table}[htb]
\begin{center}
\begin{tabular}{ccccccccc}
Model & $\ac^\text{new}$ & $\ac^\text{new} \times \sqrt \sigma$ & $y_C$ & $A_\text{cen}^\text{new}$ & $A_\text{cen}^\text{new} \times \sqrt \sigma$ & $A_\text{fwd}^\text{new}$ & $A_\text{fwd}^\text{new} \times \sqrt \sigma$\\
$Z'$ & 0.0403 & 13.0 fb$^{1/2}$
  & \begin{tabular}{c}1.5 \\ 0.7\end{tabular}
  & \begin{tabular}{c}-0.0207 \\ -0.0267\end{tabular}
  & \begin{tabular}{c}-8.6 fb$^{1/2}$ \\ -8.4 fb$^{1/2}$\end{tabular} 
  & \begin{tabular}{c}0.107 \\ 0.0239 \end{tabular}
  & \begin{tabular}{c}19.6 fb$^{1/2}$ \\ 7.9 fb$^{1/2}$ \end{tabular}
  \\[4mm]
$W'$ & 0.0536 & 18.0 fb$^{1/2}$
  & \begin{tabular}{c}1.5 \\ 0.7\end{tabular}
  & \begin{tabular}{c}-0.0249 \\ -0.0360 \end{tabular}
  & \begin{tabular}{c}-10.8 fb$^{1/2}$ \\ -11.6 fb$^{1/2}$ \end{tabular}
  & \begin{tabular}{c}0.119 \\ 0.0304 \end{tabular}
  & \begin{tabular}{c}23.5 fb$^{1/2}$ \\ 10.6 fb$^{1/2}$ \end{tabular}
  \\[4mm]
$\mathcal{G}_\mu$ & 0.0433 & 14.2 fb$^{1/2}$
  & \begin{tabular}{c}1.5 \\ 0.7\end{tabular}
  & \begin{tabular}{c}-0.0142 \\ -0.0264 \end{tabular}
  & \begin{tabular}{c}-6.0 fb$^{1/2}$ \\ -8.4 fb$^{1/2}$ \end{tabular}
  & \begin{tabular}{c}0.0800 \\ 0.0241 \end{tabular}
  & \begin{tabular}{c}14.3 fb$^{1/2}$ \\ 8.1 fb$^{1/2}$ \end{tabular}
  \\[4mm]
$\phi$ & 0.0223 & 7.2 fb$^{1/2}$ 
  & \begin{tabular}{c}1.5 \\ 0.7\end{tabular}
  & \begin{tabular}{c}-0.0071 \\ -0.0114 \end{tabular}
  & \begin{tabular}{c}-3.0 fb$^{1/2}$ \\ -3.6 fb$^{1/2}$ \end{tabular}
  & \begin{tabular}{c}0.0401 \\ 0.0104 \end{tabular}
  & \begin{tabular}{c}7.2 fb$^{1/2}$ \\ 3.4 fb$^{1/2}$ \end{tabular}
  \\[4mm]
$\omf$ & 0.0248 & 8.3 fb$^{1/2}$
  & \begin{tabular}{c}1.5 \\ 0.7\end{tabular}
  & \begin{tabular}{c}-0.0085 \\ -0.0139 \end{tabular}
  & \begin{tabular}{c}-3.7 fb$^{1/2}$ \\ -4.5 fb$^{1/2}$ \end{tabular}
  & \begin{tabular}{c}0.0480 \\ 0.0126 \end{tabular}
  & \begin{tabular}{c}8.8 fb$^{1/2}$ \\ 4.3 fb$^{1/2}$ \end{tabular}
  \\[4mm]
$\OMf$ & 0.0185 & 6.1 fb$^{1/2}$
  & \begin{tabular}{c}1.5 \\ 0.7\end{tabular}
  & \begin{tabular}{c}-0.0060 \\ -0.0110 \end{tabular}
  & \begin{tabular}{c}-2.6 fb$^{1/2}$ \\ -3.5 fb$^{1/2}$ \end{tabular}
  & \begin{tabular}{c}0.0331 \\ 0.0099 \end{tabular}
  & \begin{tabular}{c}6.1 fb$^{1/2}$ \\ 3.4 fb$^{1/2}$ \end{tabular}
\end{tabular}
\caption{Charge asymmetries corresponding to a benchmark point (see Table~\ref{tab:bench}) selected for each model.}
\label{tab:AC}
\end{center}
\end{table}
From this comparison we can see that $A_\text{fwd}$ offers some (statistical) improvement over $\ac$, but only for the models where a very light new particle is exchanged in the $t$ channel. In the remaining cases, the statistical sensitivities of these two observables are very similar. The actual sensitivity will of course depend on the systematic uncertainties, which are expected smaller with the definition in Eq.~(\ref{ec:ACMS}). On the other hand, the numbers show that the central asymmetry $A_\text{cen}$ has significantly lower sensitivity in all six models. This is not surprising, as the charge asymmetries are mostly generated at large rapidities in all the models (in agreement with the CDF results).

\begin{figure}[p]
\begin{center}
\begin{tabular}{ccc}
\epsfig{file=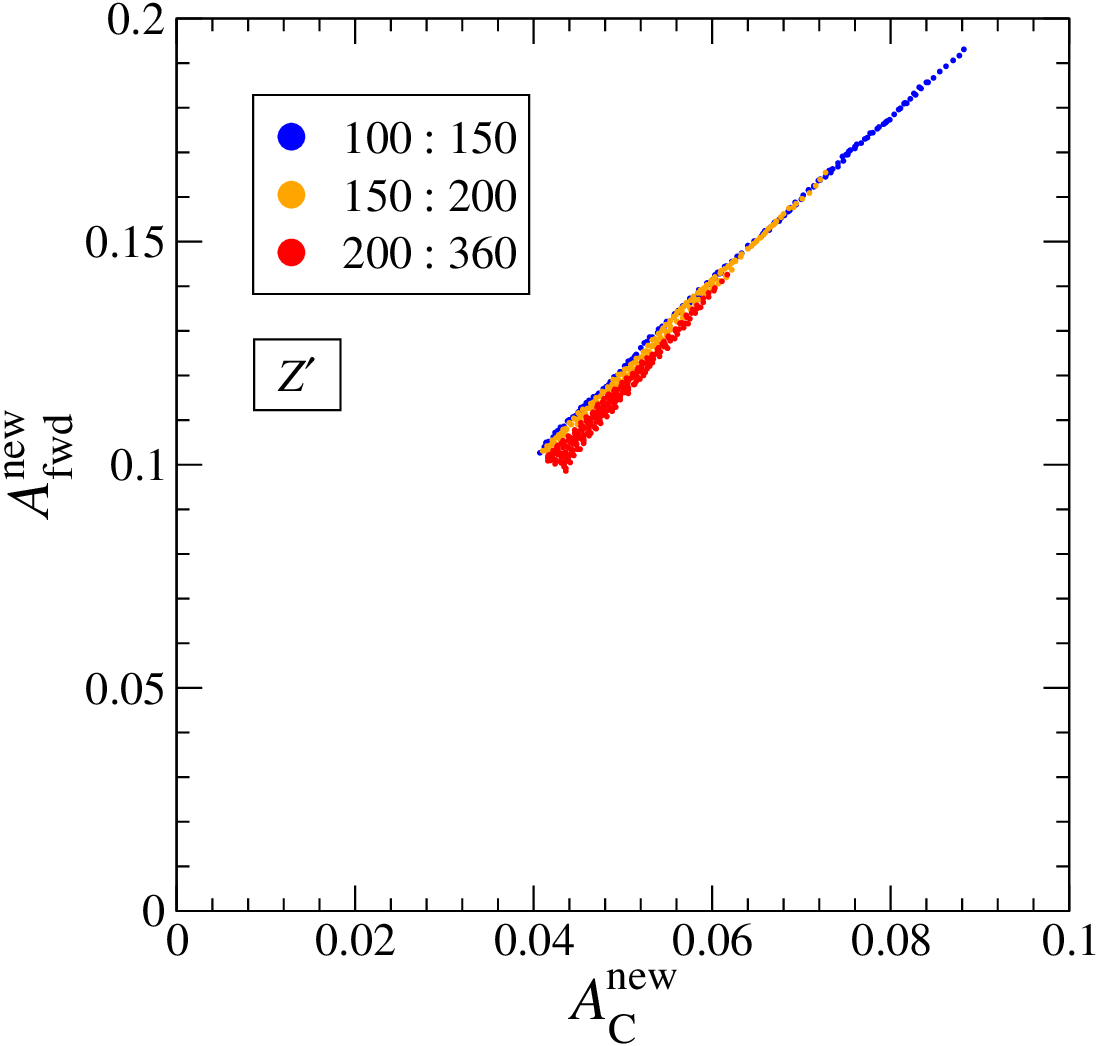,width=6.7cm,clip=} & \quad\quad &
\epsfig{file=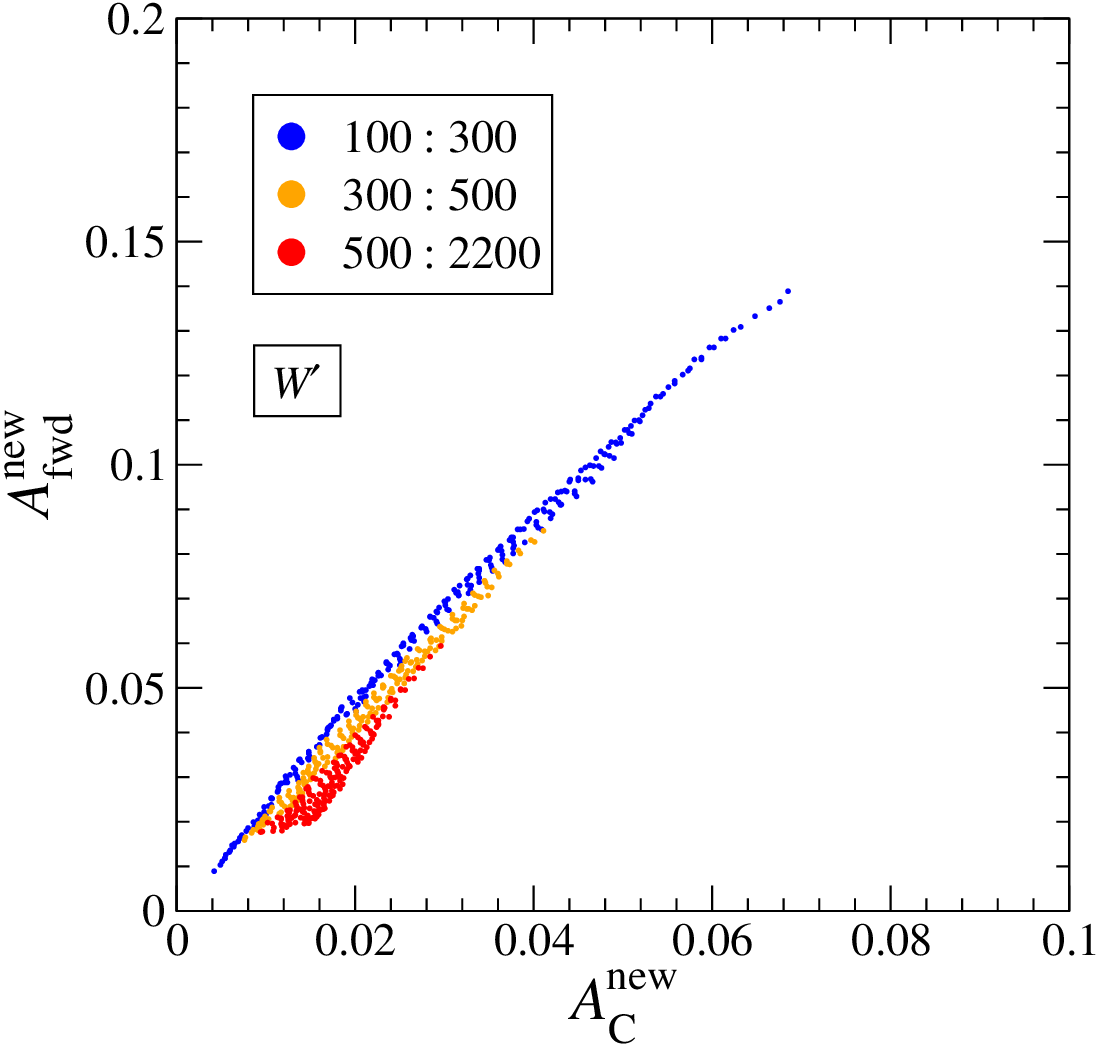,width=6.7cm,clip=} \\[2mm]
\epsfig{file=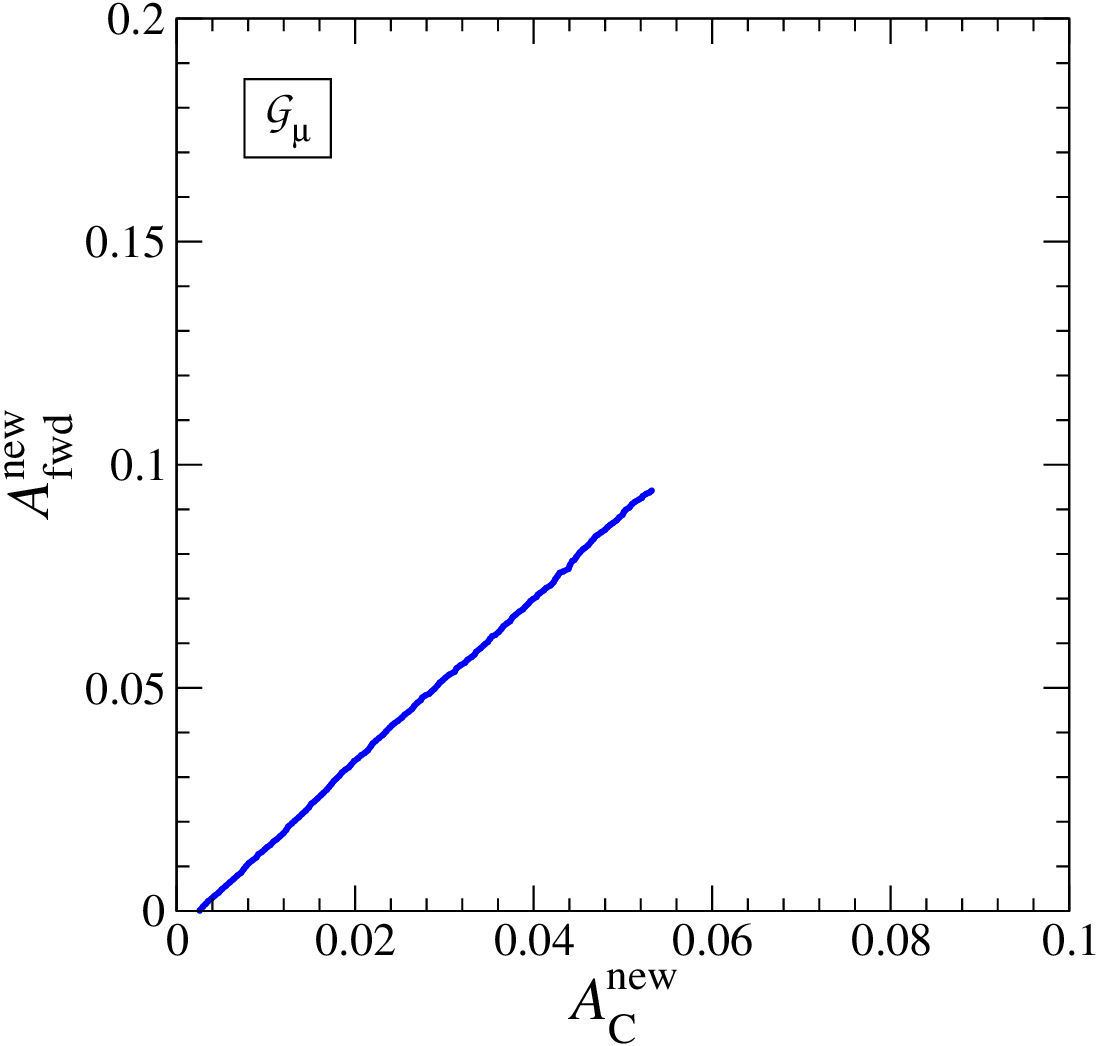,width=6.7cm,clip=} & &
\epsfig{file=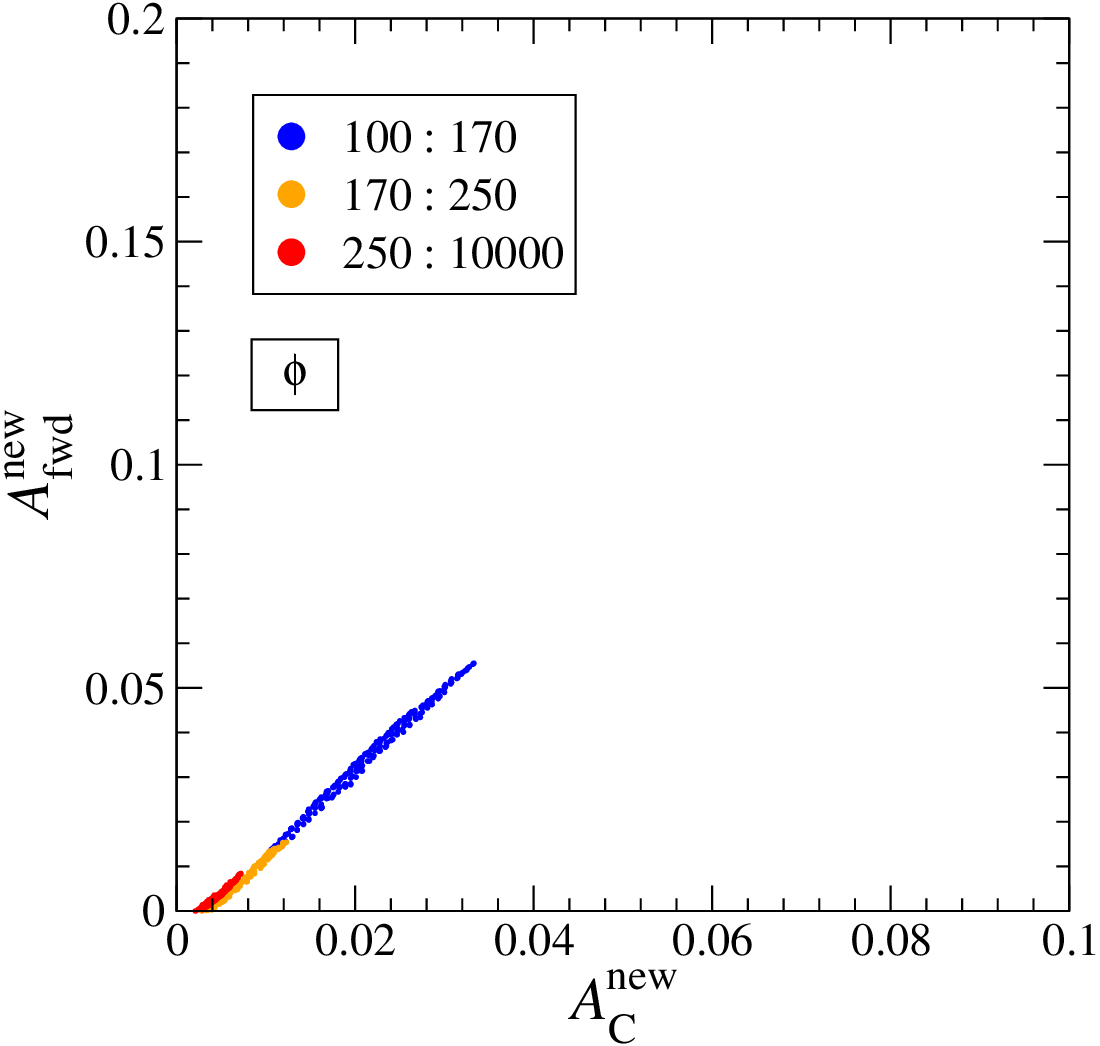,width=6.7cm,clip=} \\[2mm]
\epsfig{file=Figs/L0LF-om.eps,width=6.7cm,clip=} & &
\epsfig{file=Figs/L0LF-Om.eps,width=6.7cm,clip=}
\end{tabular}
\caption{Allowed regions for the new physics contributions to the charge asymmetries $\ac$ and $A_\text{fwd}$ (see the text). The mass ranges in the legends are in GeV.}
\label{fig:L0LF}
\end{center}
\end{figure}

\begin{figure}[p]
\begin{center}
\begin{tabular}{ccc}
\epsfig{file=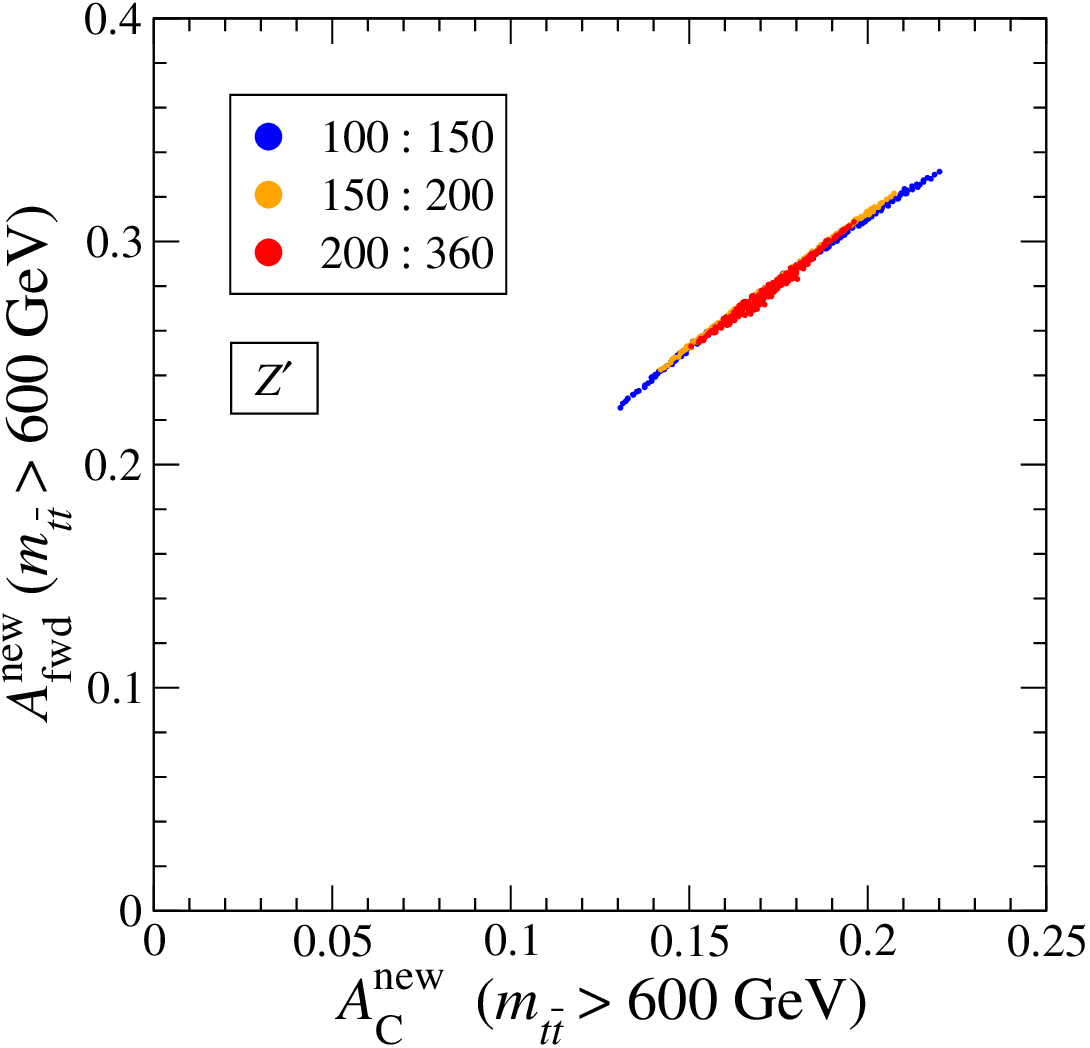,width=6.7cm,clip=} & \quad\quad &
\epsfig{file=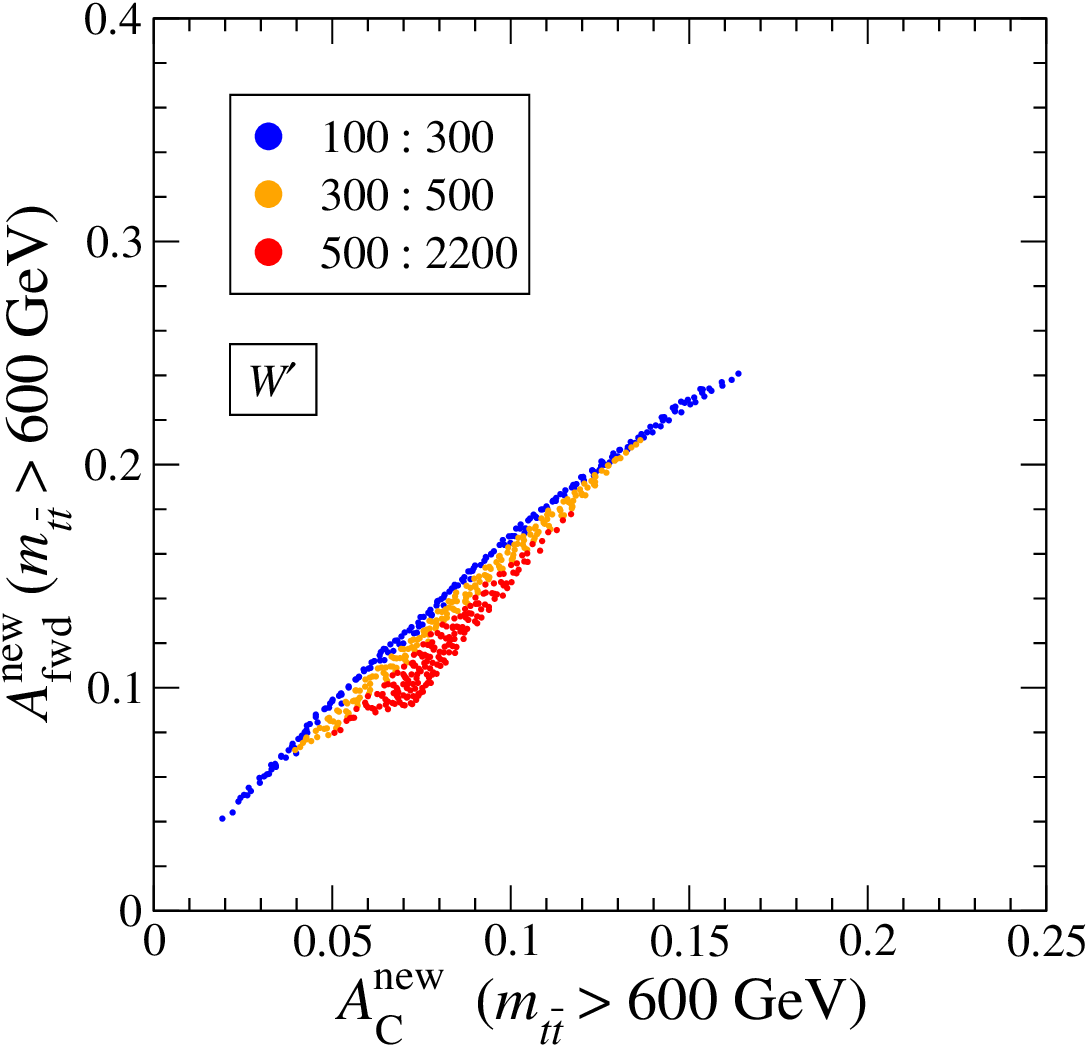,width=6.7cm,clip=} \\[2mm]
\epsfig{file=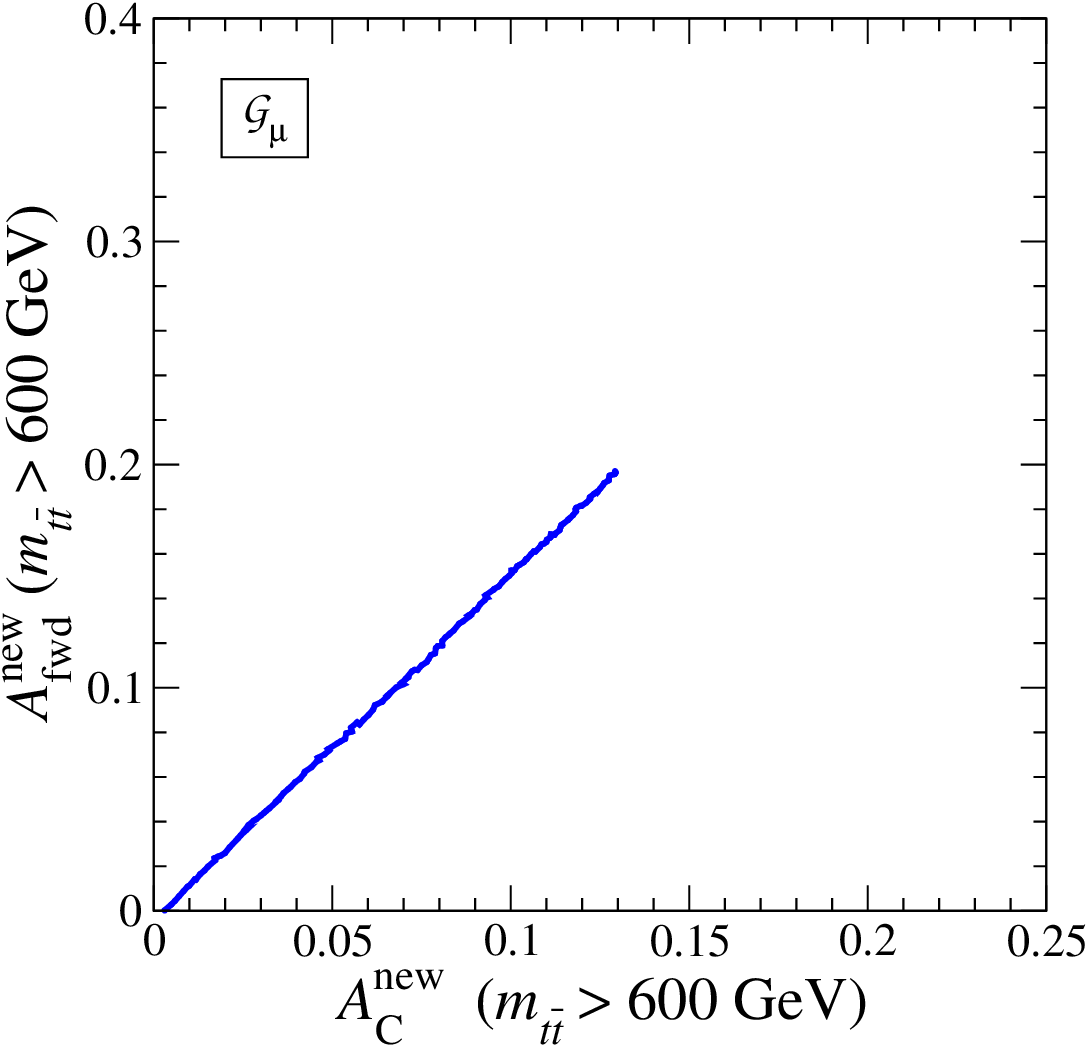,width=6.7cm,clip=} & &
\epsfig{file=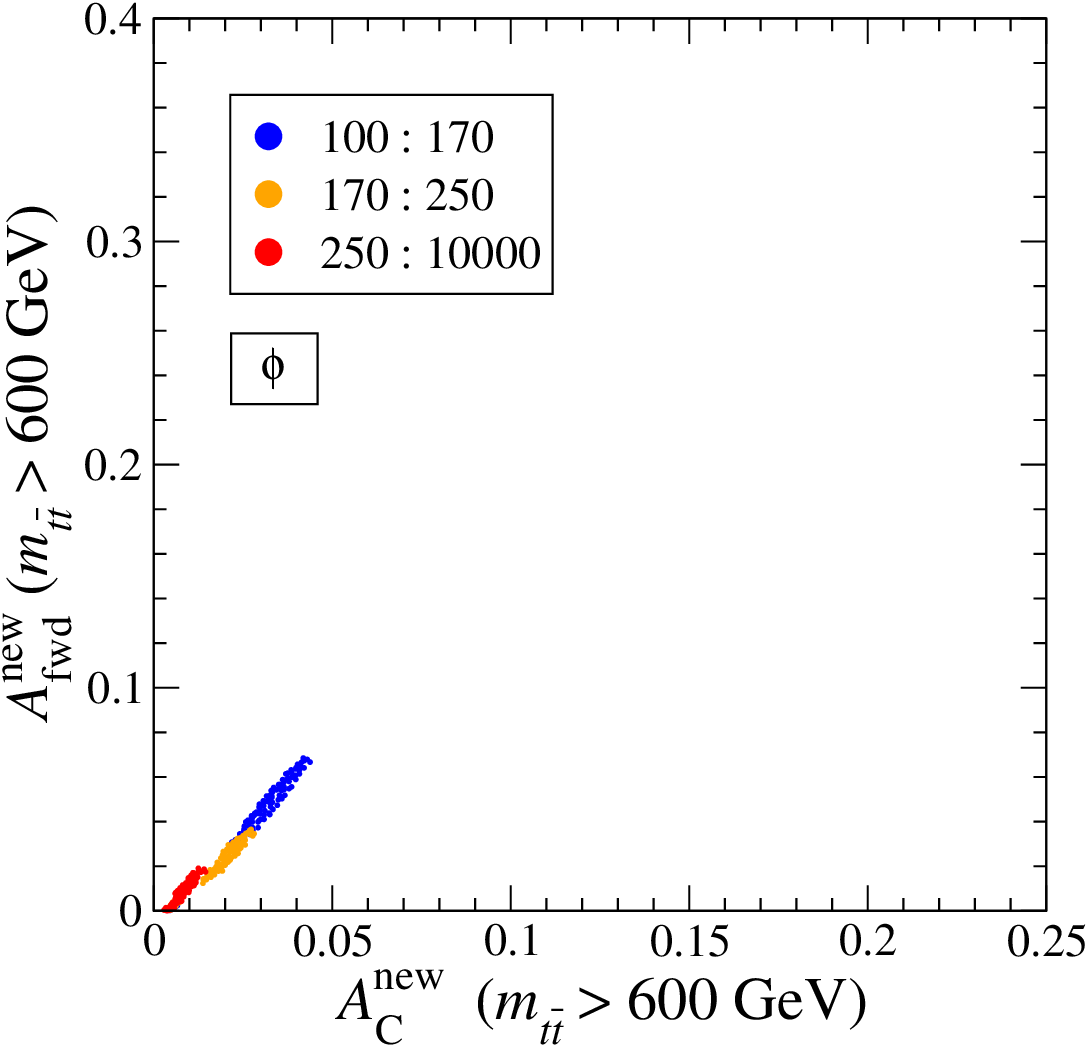,width=6.7cm,clip=} \\[2mm]
\epsfig{file=Figs/L1LF-om.eps,width=6.7cm,clip=} & &
\epsfig{file=Figs/L1LF-Om.eps,width=6.7cm,clip=}
\end{tabular}
\caption{Allowed regions for the new physics contributions to the high-mass charge asymmetries $\ac$ and $A_\text{fwd}$ (see the text). The mass ranges in the legends are in GeV.}
\label{fig:L1LF}
\end{center}
\end{figure}

We present in Fig.~\ref{fig:L0LF} the allowed regions for new physics contributions to the charge asymmetries $\ac$ and $A_\text{fwd}$ in our six models. In Fig.~\ref{fig:L1LF}, we show the same predictions but for events with invariant masses $m_{t \bar t} > 600$ GeV. We observe that, as in the case of the benchmark points in Table~\ref{tab:AC}, $A_\text{fwd}$ is larger than $\ac$. But in all cases, except for $Z'$, $W'$, the differences are compensated by the smaller statistics. This feature persists when $m_{t \bar t} > 600$ GeV is required. We conclude that the simple definition in Eq.~(\ref{ec:ACMS}) provides a good measure of the charge asymmetry at LHC, with similar statistical sensitivity to most new physics models as other definitions. Eventually, systematic uncertainties (which require a detailed evaluation) will dominate when larger data samples are collected, and having a larger asymmetry with a smaller data sample may constitute an advantage.

\section{Conclusions}

The new measurements in top pair production at Tevatron and the LHC will throw light on the present anomalies in the FB asymmetry. New effects, with definite patterns, are typically expected in most explanations invoking new physics. The new measurements will therefore discard some models and constrain the parameter space of the surviving ones. A good agreement of the LHC observables with the SM could also be a hint of less obvious forms of new physics, assuming that the Tevatron excesses persist. This could, for instance, involve the interplay of several new particles\footnote{In most models with more than one new particle, for instance a $Z'$ plus a $W'$ boson~\cite{Barger:2010mw}, the different new particles contribute to distinct processes, so the combination does not help avoiding the LHC tail constraints.} (see, e.g, \cite{Cui:2011xy}), extremely large widths for resonances in the $s$ channel~\cite{Barcelo:2011vk,AguilarSaavedra:2011ci}, unparticle stuff~\cite{Chen:2010hm} or loop effects from new physics~\cite{Gabrielli:2011jf}. 

Here, we have studied the Tevatron FB asymmetry and charge asymmetries at the LHC, for a representative selection of models. We have focussed on the implications of the measurement of the cross section at the tail of the $t\bar{t}$ invariant mass distribution, which is especially sensitive to $q\bar{q}$ initial states. We have studied the effect on the asymmetries of requiring either a loose or a precise agreement with the SM prediction. In the latter case, the $Z'$ and $W'$ models are excluded as explanations of the Tevatron anomalies. Moreover, the allowed parameter space of most models is significantly reduced: among the simple models we study, only the axigluon and the scalar doublet are able to reproduce the CDF high-invariant-mass asymmetry within two standard deviations.

We have also analysed the dependence of the Tevatron and LHC asymmetries on the masses of the exchanged particles. This dependence, as well as the differences between different models, is intensified when the asymmetries are measured in events with high invariant masses. Consequently, we stress that a study of the charge asymmetries in different invariant mass intervals at the LHC has a great importance to discriminate among different scenarios and, eventually, measure the parameters of a given model. 

Finally, we have compared the sensitivity of different definitions of the charge asymmetries at LHC. We conclude that the simple ``inclusive'' definition in Eq.~(\ref{ec:ACMS}), which uses all rapidities, is competitive with other proposals. Even if the forward asymmetry defined in Eq.~(\ref{ec:ACAF}) may perform slightly better for models in which the forward enhancement is very pronounced ($Z'$ and $W'$), these models will be ruled out soon, as explained above, if the $t\bar{t}$ cross section at LHC agrees with the SM prediction. Nevertheless, having larger asymmetries in smaller data samples may constitute an advantage in the future when the measurements are dominated by systematic uncertainties.

{\it Note added.} After the submission of this paper a new measurement by the D0 Collaboration appeared~\cite{Collaboration:2011rq} which agrees with CDF on the inclusive value of the asymmetry but does not find such a strong enhancement for high invariant masses. The implications concerning this measurement, or future ones, can easily be drawn from the plots presented.

\section*{Acknowledgements}

This work has been partially supported by projects FPA2010-17915 (MICINN), FQM 101 and FQM 437 (Junta de Andaluc\'{\i}a) and CERN/FP/116397/2010 (FCT).

\end{document}